\documentclass[10pt]{iopart}
\begin{document}

\title{The physics of star formation}

\author{Richard B Larson}

\address{Department of Astronomy, Yale University, Box 208101, New Haven,
         CT 06520-8101, USA}
\ead{larson@astro.yale.edu}

\begin{abstract}

   Our current understanding of the physical processes of star formation is reviewed, with emphasis on processes occurring in molecular clouds like those observed nearby.  The dense cores of these clouds are predicted to undergo gravitational collapse characterized by the runaway growth of a central density peak that evolves toward a singularity.  As long as collapse can occur, rotation and magnetic fields do not change this qualitative behavior.  The result is that a very small embryonic star or protostar forms and grows by accretion at a rate that is initially high but declines with time as the surrounding envelope is depleted.  Rotation causes some of the remaining matter to form a disk around the protostar, but accretion from protostellar disks is not well understood and may be variable.  Most, and possibly all, stars form in binary or multiple systems in which gravitational interactions can play a role in redistributing angular momentum and driving episodes of disk accretion.  Variable accretion may account for some peculiarities of young stars such as flareups and jet production, and protostellar interactions in forming systems of stars will also have important implications for planet formation.  The most massive stars form in the densest environments by processes that are not yet well understood but may include violent interactions and mergers.  The formation of the most massive stars may have similarities to the formation and growth of massive black holes in very dense environments.

\end{abstract}

\section{Introduction} % 1

   Stars are the fundamental units of luminous matter in the universe, and they are responsible, directly or indirectly, for most of what we see when we observe it.  They also serve as our primary tracers of the structure and evolution of the universe and its contents.  Consequently, it is of central importance in astrophysics to understand how stars form and what determines their properties.  The generally accepted view that stars form by the gravitational condensation of diffuse matter in space is very old, indeed almost as old as the concept of universal gravitational attraction itself, having been suggested by Newton in 1692\footnote{In his first letter to Bentley, as quoted by Jeans (1929), Newton said ``It seems to me, that if the matter of our sun and planets, and all the matter of the universe, were evenly scattered throughout all the heavens, and every particle had an innate gravity towards all the rest, and the whole space throughout which this matter was scattered, was finite, the matter on the outside of this space would by its gravity tend towards all the matter on the inside, and by consequence fall down into the middle of the whole space, and there compose one great spherical mass.  But if the matter were evenly disposed throughout an infinite space, it could never convene into one mass; but some of it would convene into one mass and some into another, so as to make an infinite number of great masses, scattered great distances from one to another throughout all that infinite space.  And thus might the sun and fixed stars be formed, supposing the matter were of a lucid nature.''}.  However, it is only in the past half century that the evidence has become convincing that stars are presently forming by the condensation of diffuse interstellar matter in our Galaxy and others, and it is only in recent decades that we have begun to gain some physical understanding of how this happens.  Observations at many wavelengths, especially radio and infrared, have led to great advances in our knowledge of the subject, and the observational study of star formation is now a large and active field of research.  Extensive theoretical and computational work has also contributed increasingly to clarifying the physical processes involved.

   Star formation occurs as a result of the action of gravity on a wide range of scales, and different mechanisms may be important on different scales, depending on the forces opposing gravity.  On galactic scales, the tendency of interstellar matter to condense under gravity into star-forming clouds is counteracted by galactic tidal forces, and star formation can occur only where the gas becomes dense enough for its self-gravity to overcome these tidal forces, for example in spiral arms.  On the intermediate scales of star-forming `giant molecular clouds', turbulence and magnetic fields may be the most important effects counteracting gravity, and star formation may involve the dissipation of turbulence and magnetic fields.  On the small scales of individual prestellar cloud cores, thermal pressure becomes the most important force resisting gravity, and it sets a minimum mass that a cloud core must have to collapse under gravity to form stars.  After such a cloud core has begun to collapse, the centrifugal force associated with its angular momentum eventually becomes important and may halt its contraction, leading to the formation of a binary or multiple system of stars.  When a very small central region attains stellar density, its collapse is permanently halted by the increase of thermal pressure and an embryonic star or `protostar' forms and continues to grow in mass by accretion.  Magnetic fields may play a role in this final stage of star formation, both in mediating gas accretion and in launching the bipolar jets that typically announce the birth of a new star.

   In addition to these effects, interactions between the stars in a forming multiple system or cluster may play an important role in the star formation process.  Most, and possibly all, stars form with close companions in binary or multiple systems or clusters, and gravitational interactions between the stars and gas in these systems may cause the redistribution of angular momentum that is necessary for stars to form.  Interactions in dense environments, possibly including direct stellar collisions and mergers, may play a particularly important role in the formation of massive stars.  Such processes, instead of generating characteristic properties for forming stars, may be chaotic and create a large dispersion in the properties of stars and stellar systems.  Thus star formation processes, like most natural phenomena, probably involve a combination of regularity and randomness.

   Some outcomes of star formation processes that are particularly important to understand include the rate at which the gas in galaxies is turned into stars, and the distribution of masses with which stars are formed.  The structures of galaxies depend on the circumstances in which stars form and the rate at which they form, while the evolution of galaxies depends on the spectrum of masses with which they form, since low-mass stars are faint and evolve slowly while massive ones evolve fast and release large amounts of matter and energy that can heat and ionize the interstellar gas, enrich it with heavy elements, and possibly expel some of it into intergalactic space.  It is also important to understand the formation of binary systems because many important astrophysical processes, including the formation of various kinds of exotic objects, involve the interactions of stars in binary systems.  A further outcome of star formation that is of great interest to understand is the formation of planetary systems, which may often form as byproducts of star formation in disks of leftover circumstellar material.

   The aim of this review is to summarize our current understanding of the physical processes of star formation, with emphasis on the processes occurring on small scales in star-forming molecular clouds.  Previous reviews of the small-scale processes of star formation include those by Hayashi (1966), Larson (1973), Tohline (1982), Shu \etal (1987, 1993), Bodenheimer (1992), Hartmann (1998), and Sigalotti and Klapp (2001).  A review with emphasis on the role of turbulence has been given by Mac Low and Klessen (2003), and reviews focusing on the larger-scale aspects of star formation have been given by Tenorio-Tagle and Bodenheimer (1988), Elmegreen (1992), Larson (1992), and Kennicutt (1998).  Very useful topical reviews of many aspects of star and planet formation have been published in the {\it `Protostars and Planets'} series of volumes edited by Gehrels (1978), Black and Matthews (1985), Levy and Lunine (1993), and Mannings \etal (2000), and also in the review volumes edited by Tenorio-Tagle \etal (1992) and Lada and Kylafis (1991, 1999).

\section{Observed properties of star-forming clouds} % 2

\subsection{Sites of star formation} % 2.1

   Most of the star formation in galaxies occurs in spiral arms, which are marked primarily by their concentrations of luminous young stars and associated glowing clouds of ionized gas.  The most luminous stars have lifetimes shorter than 10~million years, or $10^{-3}$ times the age of the universe, so they must have formed very recently from the dense interstellar gas that is also concentrated in the spiral arms.  Star formation occurs also near the centers of some galaxies, including our own Milky Way galaxy, but this nuclear star formation is often obscured by interstellar dust and its existence is inferred only from the infrared radiation emitted by dust heated by the embedded young stars.  The gas from which stars form, whether in spiral arms or in galactic nuclei, is concentrated in massive and dense `molecular clouds' whose hydrogen is nearly all in molecular form.  Some nearby molecular clouds are seen as `dark clouds' against the bright background of the Milky Way because their interstellar dust absorbs the starlight from the more distant stars.

   In some nearby dark clouds many faint young stars are seen, most distinctive among which are the T~Tauri stars, whose variability, close association with the dark clouds, and relatively high luminosities for their temperatures indicate that they are extremely young and have ages of typically only about 1~million years (Herbig 1962; Cohen and Kuhi 1979).  These T~Tauri stars are the youngest known visible stars, and they are `pre-main-sequence' stars that have not yet become hot enough at their centers to burn hydrogen and begin the main-sequence phase of evolution.  Some of these young stars are embedded in particularly dense small dark clouds, which are thus the most clearly identified sites of star formation.  These clouds have been studied extensively, using radio techniques to observe the heavier molecules such as CO and infrared techniques to study the dust.  Observations of the thermal emission from the dust at far-infrared wavelengths have proven to be particularly useful for studying the structure of these small star-forming clouds; the dust is the best readily observable tracer of the mass distribution because most of the heavier molecules freeze out onto the dust grains at high densities.  Many of these small clouds are dense enough for gravity to hold them together against pressure and cause them to collapse into stars, strengthening their identification as stellar birth sites (Ward-Thompson 2002).

   The smaller and more isolated dark clouds have received the most attention because they are easiest to study and model, but most stars actually form in larger groups and clusters and in larger and more complex concentrations of molecular gas.  There is no clear demarcation between  molecular concentrations of different size, and no generally accepted terminology for them, but the terms `cloud', `clump', and `core' have all often been used, generally with reference to structures of decreasing size.  In this review, the term `clump' will be used to denote any region of enhanced density in a larger cloud, while the term `core' will be used to denote a particularly dense self-gravitating clump that might collapse to form a star or group of stars.  The term `globule' has also been used to describe some compact and isolated dark clouds (Leung 1985) whose importance as stellar birth sites was advocated by Bok (Bok 1948; Bok \etal 1971) before their role in star formation was established; Bok's enthusiastic advocacy of these globules as sites of star formation has since been vindicated, and we now know that some of them are indeed forming stars.

\subsection{Structure of molecular clouds} % 2.2

   Surveys of the molecular gas in galaxies show that it is typically concentrated in large complexes or spiral arm segments that have sizes up to a kiloparsec and masses up to $10^7$ solar masses (M$_\odot$) (Solomon and Sanders 1985; Elmegreen 1985, 1993).  These complexes may contain several giant molecular clouds (GMCs) with sizes up to 100 parsecs (pc) and masses up to $10^6$~M$_\odot$, and these GMCs in turn contain much smaller scale structure that may be filamentary or clumpy on a wide range of scales (Blitz 1993; Blitz and Williams 1999; Williams \etal 2000).  The substructures found in GMCs range from massive clumps with sizes of a several parsecs and masses of thousands of solar masses, which may form entire clusters of stars, to small dense cloud cores with sizes of the order of 0.1~pc and masses of the order of 1~M$_\odot$, which may form individual stars or small multiple systems (Myers 1985, 1999; Cernicharo 1991; Lada \etal 1993; Williams \etal 2000; Andr\'e \etal 2000; Visser \etal 2002).  The internal structure of molecular clouds is partly hierarchical, consisting of smaller subunits within larger ones, and fractal models may approximate some aspects of this structure (Scalo 1985, 1990; Larson 1995; Simon 1997; Elmegreen 1997, 1999; Stutzki \etal 1998; Elmegreen \etal 2000).  In particular, the irregular boundaries of molecular clouds have fractal-like shapes resembling those of surfaces in turbulent flows, and this suggests that the shapes of molecular clouds may be created by turbulence (Falgarone and Phillips 1991; Falgarone \etal 1991).

   Molecular clouds are the densest parts of the interstellar medium, and they are surrounded by less dense envelopes of atomic gas.  The abundance of molecules increases with density because hydrogen molecules form on the surfaces of dust grains and the rate of this process increases with increasing density.  In addition, the survival of the molecules requires that the clouds have a sufficient opacity due to dust to shield the molecules from ultraviolet radiation capable of dissociating them, and this means that molecular clouds must have a column density of at least 20 M$_\odot$~pc$^{-2}$ (Elmegreen 1985, 1993).  Most molecular clouds have column densities much higher than this and are therefore quite opaque, a typical column density being of the order of 100 M$_\odot$~pc$^{-2}$.  Because of the high densities of molecular clouds, the rate at which they are cooled by collisionally excited atomic and molecular emission processes is high, and because of their high opacity, the rate at which they are heated by external radiation is low; the result is that molecular clouds are very cold and have typical temperatures of only about 10--20~K.  Higher temperatures of up to 100~K or more may exist locally in regions heated by luminous newly formed stars.  In typical molecular clouds, cooling is due mostly to the emission of far-infrared radiation from molecules such as CO, which is usually the most important coolant (McKee \etal 1982; Gilden 1984).  However, in the densest collapsing cloud cores the gas becomes thermally coupled to the dust, which then controls the temperature by its strongly temperature-dependent thermal emission, maintaining a low and almost constant temperature of about 10~K over a wide range of densities (Hayashi and Nakano 1965; Hayashi 1966; Larson 1973, 1985; Tohline 1982; Masunaga and Inutsuka 2000).  This low and nearly constant temperature is an important feature of the star formation process, and is what makes possible the collapse of prestellar cloud cores with masses as small as one solar mass or less.

   The gas densities in molecular clouds vary over many orders of magnitude: the average density of an entire GMC may be of the order of 20 H$_2$~molecules per cm$^3$, while the larger clumps within it may have average densities of the order of $10^3$~H$_2$~cm$^{-3}$ and the small pre-stellar cloud cores may have densities of $10^5$ H$_2$~cm$^{-3}$ or more.  At the high densities and low temperatures characteristic of molecular clouds, self-gravity is important and it dominates over thermal pressure by a large factor, except in the smallest clumps.  If thermal pressure were the only force opposing gravity, molecular clouds might then be expected to collapse rapidly and efficiently into stars.  Most molecular clouds are indeed observed to be forming stars, but they do so only very inefficiently, typically turning only a few percent of their mass into stars before being dispersed.  The fact that molecular clouds do not quickly turn most of their mass into stars, despite the strong dominance of gravity over thermal pressure, has long been considered problematic, and has led to the widely held view that additional effects such as magnetic fields or turbulence support these clouds in near-equilibrium against gravity and prevent a rapid collapse.

   However, the observed structure of molecular clouds does not resemble any kind of equilibrium configuration, but instead is highly irregular and filamentary and often even windblown in appearance, suggesting that these clouds are actually dynamic and rapidly changing structures, just like terrestrial clouds.  The complex structure of molecular clouds is important to understand because it may influence or determine many of the properties with which stars and systems of stars are formed.  For example, stars often appear to form in a hierarchical arrangement consisting of smaller groupings within larger ones, and this may reflect the hierarchical and perhaps fractal-like structure of star-forming clouds (Gomez \etal 1993; Larson 1995; Testi \etal 2000; Elmegreen \etal 2000).  Stars may also derive their masses directly from those of the prestellar cloud cores, as is suggested by the fact that the distribution of masses or `initial mass function' (IMF) with which stars are formed appears to resemble the distribution of masses of the prestellar cores in molecular clouds (Motte \etal 1998; Testi and Sargent 1998; Luhman and Rieke 1999; Motte and Andr\'e 2001a,b).

\subsection{The role of turbulence and magnetic fields} % 2.3

   In addition to their irregular shapes, molecular clouds have complex internal motions, as is indicated by the broad and often complex profiles of their molecular emission lines.  In all but the smallest clumps, these motions are supersonic, with velocities that significantly exceed the sound speed of 0.2~km~s$^{-1}$ typical for dark clouds (Larson 1981; Myers 1983; Dickman 1985).  The broad line profiles appear to reflect mostly small-scale random motions rather than large-scale systematic motions such as cloud rotation, since observed large-scale motions are usually too small to contribute much to the line widths.  The internal random motions in molecular clouds are often referred to as `turbulence,' even though their detailed nature remains unclear and they may not closely resemble classical eddy turbulence.  Nevertheless the existence of some kind of hierarchy of turbulent motions is suggested by the fact that the velocity dispersion inferred from the line width increases systematically with region size in a way that resembles the classical Kolmogoroff law (Larson 1979, 1981; Myers 1983, 1985; Scalo 1987; Myers and Goodman 1988; Falgarone \etal 1992).  Supersonic turbulence may play an important role in structuring molecular clouds, since supersonic motions can generate shocks that produce large density fluctuations.  Much effort has therefore been devoted to studying the internal turbulent motions in molecular clouds, and `size-linewidth relations' have been found in many studies, albeit with considerable variability and scatter (e.g., Goodman \etal 1998; Myers 1999).

   Studies that include motions on larger scales suggest that a similar correlation between velocity dispersion and region size extends up to galactic scales, and this suggests that the turbulent motions in molecular clouds are part of a larger-scale hierarchy of interstellar turbulent motions (Larson 1979; Goldman 2000).  Molecular clouds must then represent condensations in a generally turbulent interstellar medium, and their structure and dynamics must constitute part of the structure and dynamics of the medium as a whole.  The origin of the observed large-scale interstellar motions is not yet fully understood, but many different sources almost certainly contribute to them, and the sources and properties of the interstellar turbulence may vary from place to place.  Some likely sources include gravitationally driven motions on large scales and stellar feedback effects such as ionization, winds, and supernova explosions on smaller scales.  Different star-forming regions do indeed show different levels of turbulence; for example there is a higher level of turbulence in the Orion region, where stars of all masses are forming vigorously in two large GMCs, than in the smaller and more quiescent Taurus clouds, which are forming only less massive stars (Larson 1981).  In many cases, self-gravity is roughly balanced by the turbulent motions in  molecular clouds, and this suggests that gravity and turbulence are equally important in controlling the structure and evolution of these clouds.

   In addition to being turbulent, molecular clouds are also significantly magnetized, and magnetic fields can also be important for the dynamics and evolution of these clouds (Heiles \etal 1993; McKee \etal 1993).  If molecular clouds are sufficiently strongly magnetized, their internal motions might be predominantly wavelike, consisting basically of magnetohydrodynamic (MHD) waves such as Alfv\'en waves; since Alfv\'en waves involve only transverse and non-compressional motions, they might be expected to dissipate more slowly than purely hydrodynamic supersonic turbulence (Arons and Max 1975).  Wavelike `MHD turbulence' might then provide a source of pressure that can supplement thermal pressure and help to support molecular clouds for a significant time against gravity (Myers and Goodman 1988; McKee \etal 1993; McKee and Zweibel 1995).  If molecular clouds can indeed be supported against gravity for a long time by a combination of static magnetic fields and slowly dissipating MHD turbulence, this might provide some justification for models that treat these clouds as long-lived quasi-equilibrium structures (Shu \etal 1987, 1993, 1999; McKee \etal 1993; McKee 1999).

   It is difficult to test models that assume important magnetic cloud support because of the paucity of accurate measurements of field strengths; direct measurements using the Zeeman effect are difficult and in most cases have yielded only upper limits (Heiles \etal 1993).  A compilation of results by Crutcher (1999) suggests that static magnetic fields are not sufficient by themselves to counterbalance gravity, but that there may be a rough equipartition between the magnetic energy and the turbulent kinetic energy in molecular clouds, in which case a combination of static magnetic fields and MHD turbulence might together be able to support these clouds against gravity.  Bourke \etal (2001) find, with further data, that the measured magnetic field is not sufficient to balance gravity if the clouds studied are spherical but could be sufficient if these clouds are flattened and the undetected fields are close to their upper limits.  If static magnetic fields were to be important in supporting molecular clouds against gravity, however, one might expect to see alignments between cloud structures and the magnetic field direction inferred from polarization studies, but efforts to find such alignments have yielded ambiguous results and often do not show the expected alignments (Goodman \etal 1990).

   Progress in understanding the role of MHD turbulence in molecular clouds has in the meantime come from numerical simulations (Ostriker \etal 1999; Vazquez-Semadeni \etal 2000; Nordlund and Padoan 2003; Mac Low and Klessen 2003; Ballesteros-Paredes 2003).  These simulations show that, even if magnetic fields are important and the turbulence is predominantly wavelike, any MHD wave motions decay in a time comparable to the dynamical or crossing time of a cloud, defined as the cloud size divided by a typical turbulent velocity (Mac Low \etal 1998; Stone \etal 1998; Padoan and Nordlund 1999).  This rapid wave dissipation occurs because, even if all of the wave energy is initially in transverse motions, motions along the field lines are immediately generated and they produce shocks that soon dissipate the wave energy.  Thus, gravity cannot be balanced for long by any kind of turbulence unless the turbulence is continually regenerated by a suitable energy source.  But fine tuning is then needed, and sufficient turbulence must be generated on all scales to maintain a cloud in equilibrium without disrupting it (Mac Low and Klessen 2003).  Models of this kind are constrained by the fact that the heating associated with the dissipation of internally generated turbulence may produce temperatures higher than are observed (Basu and Murali 2001).

   If molecular clouds are in fact transient rather than quasi-equilibrium structures, as is suggested by the evidence on their lifetimes (Larson 1994; see below), this removes much of the motivation for postulating long-term magnetic support.  Magnetic fields may still have important consequences for star formation, but they may exert their most important effects during the early phases of cloud evolution when the fields are still strongly coupled to the gas and can damp rotational motions, thus helping to solve the `angular momentum problem' of star formation (Mouschovias 1977, 1991; see Secion~6).  During the later high-density stages of collapse, the gas is expected to decouple from the magnetic field by ambipolar diffusion, and the magnetic field then becomes dynamically unimportant until a much later stage when stellar conditions are approached at the center (Basu and Mouschovias 1994; Mouschovias and Ciolek 1999).

\subsection{Cloud evolution and lifetimes} % 2.4

   Evidence concerning the lifetimes and evolution of molecular clouds is provided by the ages of the associated newly formed stars and star clusters (Blaauw 1964, 1991; Larson 1981, 1994; Elmegreen 2000; Andr\'e \etal 2000; Hartmann 2001, 2003).  Very few giant molecular clouds are known that are not forming stars, and the most massive and dense ones all contain newly formed stars.  This means that there cannot be any significant `dead time' between the formation of a massive dense molecular cloud and the onset of star formation in it; in particular, there cannot be any long period of slow quasi-static evolution before stars begin to form.  Molecular  clouds also cannot survive for long after beginning to make stars, since the age span of the associated young stars and clusters is never more than about 10 million years, about the dynamical or crossing time of a large GMC, and since stars and clusters older than 10~Myr no longer have any associated molecular gas (Leisawitz \etal 1989).  Smaller clouds have even shorter stellar age spans that in all cases are comparable to their crossing times (Elmegreen 2000); in the Taurus clouds, for example, most of the stars have formed just in the past few million years (Hartmann \etal 2001; Palla and Stahler 2002; Hartmann 2003).  On the smallest scales, prestellar cloud cores have yet shorter estimated lifetimes that are only a few hundred thousand years for the smallest and densest cores (Andr\'e \etal 2000).  Star formation is, therefore, evidently a fast process, and it always occurs in a time comparable to the crossing time of the associated molecular cloud or core.  After that, a star-forming cloud must soon be destroyed or become no longer recognizable, perhaps being destroyed by stellar feedback effects such as ionization (Tenorio-Tagle 1979; Larson 1988; Franco \etal 1994; Matzner 2002) or being dispersed or restructured by larger-scale interstellar flows.

   Studies of the chemistry of molecular clouds also suggest young ages and short lifetimes.  As discussed by Stahler (1984), Prasad \etal (1987), Herbst (1994), and van Dishoeck and Blake (1998), molecular clouds and cloud cores appear to be chemically relatively unevolved, since the observed abundances of various molecules are often far from those expected to prevail in chemical equilibrium and resemble instead those predicted to occur at an early stage of cloud evolution less than 1~Myr after their formation.  This again suggests that molecular clouds are quite young, or at least that they have undergone recent chemical reprocessing by some major restructuring event; in the latter case the observed structures must still be of recent origin.  The fact that most of the molecules in dense molecular clouds have not yet frozen out on the dust grains, as would be expected if these clouds are stable long-lived objects, also suggests that molecular clouds or cores may be relatively young, of the order of 1~Myr or less.

   The evidence therefore suggests that molecular clouds are transient structures that form, evolve, and disperse rapidly, all in a time comparable to the crossing time of their internal turbulent motions (Larson 1994; Elmegreen 2000; Hartmann 2003).  Numerical simulations of turbulence in molecular clouds also do not support the possibility that these clouds can be supported against gravity for a long time in a near-equilibrium state and suggest that the observed structures are quite transient (Mac Low and Klessen 2003; Ballesteros-Paredes 2003).  This is consistent with the irregular and often windblown appearances of molecular clouds.  If molecular clouds are indeed transient structures, they must be assembled rapidly by larger-scale motions in the interstellar medium (Ballesteros-Paredes \etal 1999).  The processes that may be involved in the formation and destruction of molecular clouds have been reviewed by Larson (1988) and Elmegreen (1993), and the processes that may be responsible for the rapid formation of stars in them during their brief existence are discussed in Sections 3 and~4 below.

\section{Fragmentation of star-forming clouds} % 3

   The youngest stars are associated with the denser parts of molecular clouds, and especially with the densest cloud cores that appear to be the direct progenitors of stars and stellar groupings (Myers 1985, 1999; Cernicharo 1991; Lada \etal 1993; Williams \etal 2000; Williams and Myers 2000; Andr\'e \etal 2000).  What mechanisms might be responsible for generating the observed clumpy structure of molecular clouds, often consisting of a hierarchy of clumps of various sizes?  Two basic types of processes could be involved: (1) The observed dense clumps and cloud cores might originate from small density fluctuations in molecular clouds that are amplified by their self-gravity; such a gravitational fragmentation process might in principle generate a hierarchy of progressively smaller and denser clumps. (2) Alternatively, the observed clumpy structure might be generated by supersonic turbulent motions that compress the gas in shocks; a hierarchy of compressed regions or clumps might then be produced by a hierarchy of turbulent motions.  Almost certainly, both gravity and turbulence play important roles in fragmenting molecular clouds into the observed dense star-forming clumps, but it will be useful first to consider their effects separately.

\subsection{Gravitational instability} % 3.1

   The classical view, dating back to the speculations of Newton and developed further by Jeans (1902, 1929), Hoyle (1953), and Hunter (1964), is that star formation begins with small density fluctuations in an initially nearly uniform medium that are amplified by gravity in a process called `gravitational instability.'  Jeans studied the growth of plane-wave density perturbations in an infinite uniform medium that has a finite pressure but no rotation, magnetic fields, or turbulence, and he showed that short-wavelength perturbations are pressure-dominated and propagate as sound waves, while perturbations whose wavelength exceeds a critical value called the `Jeans length' are gravity-dominated and do not propagate but grow exponentially.  For an isothermal medium with a uniform density $\rho$ and a constant temperature $T$ that is fixed by radiative processes, the Jeans length $\lambda_{\rm J}$ can be expressed in terms of the density $\rho$ and the isothermal sound speed $c = (kT/m)^{1/2}$, where $m$ is the average particle mass:
\begin{equation}
               \lambda_{\rm J} ~=~ \pi^{1/2}c(G\rho)^{-1/2}.        % (1)  
\end{equation}
Assuming that the density fluctuations or clumps from which stars form have similar dimensions in all three coordinates, a corresponding minimum mass for gravitationally unstable density fluctuations can be estimated; this `Jeans mass' $M_{\rm J}$, usually defined as $\rho \lambda_{\rm J}^3$, is
\begin{equation}
                 M_{\rm J} ~=~ 5.57\,c^3\!/G^{3/2}\rho^{1/2}          % (2)
\end{equation}
(Spitzer 1978).  If spherical rather than plane-wave density perturbations are assumed and if the Jeans mass is defined as the mass in the contracting region inside the first minimum of the spherical eigenfunction $r^{-1}\sin kr$, the result differs from equation (2) only in having a numerical coefficient of 8.53 instead of 5.57 (Larson 1985).

   As has often been noted, the relevance of the Jeans analysis is unclear because it is mathematically inconsistent, neglecting the collapse of the background medium which may overwhelm the collapse of individual density fluctuations.  This inconsistency has sometimes caused the Jeans analysis to be called the `Jeans swindle' (e.g., Binney and Tremaine 1987).  Rigorous stability analyses can, however, be made for a variety of equilibrium configurations that do not undergo an overall collapse like the infinite uniform medium assumed by Jeans.  One such configuration is an infinite plane-parallel layer in vertical hydrostatic equilibrium; perturbations in the surface density of such a layer can be shown to grow exponentially if their wavelengths exceed a critical value analogous to the Jeans length.  For an isothermal layer with a fixed sound speed $c$, this critical wavelength is
\begin{equation}
             \lambda_{\rm crit} ~=~ 2\pi H ~=~ 2\,c^2\!/G\mu         % (3)
\end{equation}
(Spitzer 1942, 1978), where $\mu$ is the surface density of the layer and $H = c^2/\pi G\mu$ is its scale height.  If perturbations with cylindrical instead of planar symmetry are assumed, the minimum unstable mass defined as the mass in the contracting region inside the first minimum of the appropriate eigenfunction, in this case a Bessel function, is
\begin{equation} 
                    M_{\rm crit} ~=~ 4.67\,c^4\!/G^2\mu              % (4)
\end{equation}
(Larson 1985).  This result can also be expressed in terms of the density $\rho_0$ in the midplane of the layer, and this yields $M_{\rm crit} = 5.86\,c^3\!/G^{3/2}\rho_0^{1/2}$, which is almost identical to the Jeans mass as given by equation (2).  In the case of a plane layer, the growth rate is maximal for perturbations whose wavelength is almost exactly twice the critical value, and it declines for longer wavelengths (Simon 1965; Elmegreen and Elmegreen 1978), allowing fluctuations of this size to collapse faster than larger regions.  These results can be generalized to non-isothermal equations of state and are only moderately sensitive to the assumed equation of state (Larson 1985); even an incompressible layer is unstable to the growth of perturbations in its surface density (Goldreich and Lynden-Bell 1965a), and the minimum unstable mass in this case, expressed in terms of the density and sound speed in the midplane of the layer, is 2.4 times that in the isothermal case.

   Another type of configuration that has been much studied is equilibrium cylinders or filaments; such configurations might be more realistic than sheets, given the often filamentary appearance of molecular clouds (Schneider and Elmegreen 1979; Hartmann 2002) and the frequent tendency of numerical simulations of cloud collapse and fragmentation to develop filamentary structure (Larson 1972a; Monaghan and Lattanzio 1981; Miyama \etal 1987; Bodenheimer \etal 2000; Klessen and Burkert 2001; Bonnell and Bate 2002; Bate \etal 2003).  The stability of incompressible cylinders was first studied by Chandrasekhar and Fermi (1953), and the stability of more realistic isothermal filaments was studied by Stodolkiewicz (1963).  The available results, including an intermediate case with a ratio of specific heats $\gamma = 2$, are again not very sensitive to the equation of state: if the minimum unstable mass is expressed in terms of the density and sound speed on the central axis of the filament, it differs from the Jeans mass given in equation (2) only by having a numerical coefficient that varies between 3.32 and 6.28 for $1 \le \gamma \le \infty$, instead of being equal to 5.57 (Larson 1985) (no stable cylindrical equilibrium is possible for $\gamma < 1$.)  Thus, the Jeans mass appears to provide a useful approximation to the minimum mass for fragmentation that is valid quite generally, regardless of the exact geometry, equation of state, or state of equilibrium of the fragmenting configuration.

   A stability analysis can also be made for an equilibrium isothermal sphere of finite size with a fixed temperature and boundary pressure; such a configuration might be relevant to star formation if prestellar cloud cores when formed are nearly in equilibrium and in approximate pressure balance with a surrounding medium.  For a fixed sound speed $c$ and boundary pressure $P$, an isothermal sphere is unstable to collapse if its radius and mass exceed the critical values
\begin{equation}
                 R_{\rm BE} ~=~ 0.48\,c^2\!/G^{1/2}P^{1/2}           % (5)
\end{equation}
\begin{equation}
                 M_{\rm BE} ~=~ 1.18\,c^4\!/G^{3/2}P^{1/2}           % (6)
\end{equation}
(Spitzer 1968).  These results were first derived independently by Bonnor (1956) and Ebert (1957), and an isothermal sphere with these critical properties is therefore often called a `Bonnor-Ebert sphere'.  These results can be related to the Jeans length and mass discussed above by noting that in an isothermal medium the pressure and density are related by $P = \rho c^2$; thus $R_{\rm BE}$ and $M_{\rm BE}$ have the same dimensional form as the Jeans length and mass given in equations (1) and (2), but with smaller numerical coefficients that reflect the fact that a Bonnor-Ebert sphere contains only matter whose density is higher than the background density, while a region one Jeans length across also includes matter of lower density that may or may not collapse along with the denser material.  Non-spherical equilibrium configurations that might be produced by the fragmentation of an isothermal filament have been studied by Curry (2000), who demonstrated the existence of a sequence of equilibria ranging from a filament with small longitudinal density fluctuations to a chain of elongated clumps; these non-spherical clumps have stability properties similar to those of Bonnor-Ebert spheres (Lombardi and Bertin 2001; Curry 2002).

\subsection{Effects of rotation and magnetic fields} % 3.2

   Many authors have studied the stability of various rotating configurations including disks that might fragment into rings or clumps; in this case the assumption of a thin disk often provides a good approximation.  The stability properties of thin disks are basically similar to those of infinite plane layers with a modification due to rotation, and the results can be generalized in a similar way to non-isothermal equations of state (Larson 1985).  In two special cases, that of a rigidly rotating disk and that of axisymmetric modes of short wavelength, a rigorous stability analysis is possible and shows that the effect of rotation is always stabilizing in that the growth rates of unstable modes and the range of unstable wavelengths are both reduced; however the wavelength of the most rapidly growing mode remains unchanged, and is again almost exactly twice the minimum unstable wavelength or Jeans length, as in the non-rotating case.  Instability can be completely suppressed if the Toomre stability parameter $Q = c\kappa/\pi G\mu$ exceeds a critical value of order unity, where $\kappa$ is the epicyclic frequency (Binney and Tremaine 1987).  For an infinitely thin isothermal disk the critical value of $Q$ is exactly unity, while for disks of finite thickness the critical value of $Q$ is somewhat smaller than unity, its exact value depending on the equation of state; the limiting case is an incompressible disk, for which the critical value of $Q$ is 0.526 (Larson 1985).

   For a differentially rotating disk, eigenfunctions of fixed form do not exist because density fluctuations are continually wound up by differential rotation, but winding-up density fluctuations can still be amplified by a finite but possibly large factor in the `swing amplification' process studied by Goldreich and Lynden-Bell (1965b) and Toomre (1981).  There is no precise stability criterion for a differentially rotating disk, but numerical simulations show that the approximate criterion $Q < 1$ still serves as a good indicator of instability; for example, Miyama \etal (1984) find that differentially rotating disks are unstable to fragmentation into clumps if $Q$ is less than about 0.75.  If fragmentation does occur, the sizes and masses of the clumps that form are similar to those that form in non-rotating layers, since rotation affects only the growth rate but not the size or mass scale of the growing density fluctuations.

   The stability of various kinds of magnetized configurations including sheets, filaments, and disks has been studied by many authors including Chandrasekhar and Fermi (1953), Pacholczyk (1963), Stodolkiewicz (1963), Nakano and Nakamura (1978), and Nakamura \etal (1993).  Nakamura (1984) studied, in addition, the stability of disks with both rotation and magnetic fields, and showed that the effects of rotation and magnetic fields on the growth rate of perturbations are additive in the linear (small amplitude) approximation.  All of these studies show that a magnetic field, like rotation, always has a stabilizing effect, and that shorter-wavelength perturbations are more strongly stabilized.  The strongest stabilizing effect occurs for a sheet or disk threaded by a perpendicular magnetic field, and in this case instability can be completely suppressed if the magnetic pressure exceeds the gas pressure in the midplane of the sheet.  However, as long as instability is not completely suppressed, the wavelengths of the growing modes are not very different from those in the non-magnetic case.  Therefore the Jeans length and mass are still approximately valid even for configurations that are partly supported by rotation or magnetic fields, as long as instability is not completely suppressed by these effects.  Thus, if gravity is strong enough to cause collapse to occur, the minimum scale on which it can occur is always approximately the Jeans scale, and structure is predicted to grow most rapidly on scales about twice the Jeans scale.

\subsection{The role of turbulence} % 3.3

   As was noted in Section 2.3, star-forming clouds have internal turbulent motions that are supersonic on all but the smallest scales, and these motions must play some role in structuring these clouds.  Interstellar turbulent motions on larger scales may even be responsible for forming molecular clouds.  Simulations of supersonic turbulence show that even if magnetic fields are important and significantly constrain the turbulent motions, shocks are still unavoidably produced by motions along the field lines, and these shocks compress the gas into structures that can be sheetlike or filamentary or clumpy (Ostriker \etal 1999, 2001; Vazquez-Semadeni \etal 2000; Klessen \etal 2000; Padoan \etal 2001; Gammie \etal 2003; Mac Low 2003).  If efficient cooling keeps the temperature nearly constant and the shocks are approximately isothermal, as is often assumed, the density of the gas behind the shock is increased by a factor equal to the square of the shock Mach number.  The Mach numbers of the observed motions in molecular clouds are typically of the order of 5 to 10, so the shocks may compress the gas by up to two orders of magnitude in density.

   The observed turbulent motions in molecular clouds become subsonic on the smallest scales, and this suggests that there may be a lower limit to the sizes of the compressed structures that can be created by turbulence.  The smallest prestellar cloud cores in fact have subsonic internal motions, and they also appear to have relatively smooth and regular structures, possibly reflecting the fact that the subsonic turbulence in them cannot produce large density fluctuations (Larson 1981; Myers 1983, 1985; Goodman \etal 1998; Padoan \etal 2001).  The sizes of the smallest star-forming units might then be determined by the scale on which the cloud turbulence becomes subsonic; the transition from supersonic to subsonic motions occurs at a scale of the order of 0.05 to 0.1 parsecs, which is approximately the size scale of the observed prestellar cores.

   The minimum scale for turbulent fragmentation determined in this way may be essentially identical to the Jeans scale in the compressed regions created by the turbulence.  The empirical correlations among region size, velocity dispersion, and density that have been found in many molecular clouds and clumps (Larson 1981; Myers 1985; Myers and Goodman 1988; Falgarone \etal 1992) suggest that there may be a typical turbulent ram pressure $\rho v^2$ of the order of $4 \times 10^{-11}$ dynes~cm$^{-2}$ that is approximately independent of region size.  If compressed regions are created with thermal pressures of this order, the Jeans length in such regions is of the order of 0.1~pc; a Bonnor-Ebert sphere with this boundary pressure and a temperature of 10~K has a diameter of 0.06~pc and a mass of 0.7~M$_\odot$, similar to the observed sizes and masses of the prestellar cores in molecular clouds.  Thus, turbulent compression with a pressure of the above order might account for much of the observed small-scale structure in molecular clouds.  Vazquez-Semadeni \etal (2000) and Ballesteros-Paredes (2003) have emphasized that structures created by turbulence are generally transient and far from equilibrium, but if some compressed structures happen to have about the Jeans size, they  might survive longer than others and show a rough balance between gravity and pressure before collapsing or being dispersed.

   The view that star-forming cloud cores are created by turbulence thus appears to provide an attractive basis for understanding how star formation is initiated in molecular clouds (Mac Low and Klessen 2003), especially if there is a characteristic turbulent pressure determined by the large-scale properties of the interstellar medium, as suggested by Larson (1996).  However, it remains unclear to what extent the observed turbulent motions in molecular clouds may be a cause and to what extent they may be a consequence of gravitational collapse and fragmentation, which can involve very complex and even chaotic dynamics (Section~6).  The results of simulations of collapsing and fragmenting clouds do not appear to be very sensitive to the way in which turbulence is introduced, or even to whether turbulence is initially present at all; the scale of fragmentation always seems to be similar to the initial Jeans mass, although fragment masses may be somewhat reduced by compression occurring during the collapse (Larson 1978; Klessen 2001b; Bate \etal 2002a, 2003; Bonnell and Bate 2002).  Thus it could be that turbulence, like rotation and magnetic fields, plays more of a modulating than a controlling role in star formation, perhaps influencing details like the statistical properties and the spatial distribution and of young stars and stellar systems.

\section{Collapse of prestellar cloud cores} % 4

\subsection{Initial conditions} % 4.1

   The outcome of the collapse of a prestellar cloud core depends on the initial conditions and how the collapse is initiated.  We have seen that star-forming cores are created by complex processes of cloud dynamics that are not yet fully understood, so we cannot yet specify precisely how they begin their collapse.  However, two kinds of models have been widely studied that illustrate two limiting possibilities for how the collapse of a spherical cloud core might be initiated.  One possibility, suggested by the stability analyses and fragmentation simulations discussed above, is that collapse begins with an unstable or marginally stable clump of gas in which gravity gains the upper hand over thermal pressure and causes a runaway collapse to occur (Hayashi 1966).  Many calculations of the collapse of prestellar cloud cores have for example assumed that the initial state is similar to a Bonnor-Ebert sphere that slightly exceeds the stability threshold, a model that might have some theoretical plausibility and that appears to approximate the observed structures of many prestellar cloud cores (Ward-Thompson 2002).

   A different type of model is based on the assumption that prestellar cores are initially magnetically supported and condense gradually by ambipolar diffusion, whereby the gas contracts slowly across the field lines (Shu 1977; Shu \etal 1987).  Shu (1977) argued that such a quasi-static contraction process causes the core to become increasingly centrally condensed and increasingly supported by thermal pressure, eventually becoming a singular isothermal sphere (SIS) with no magnetic support and with a density distribution given by $\rho = c^2\!/2\pi Gr^2$.  Such a configuration is unstable and probably unattainable by any real physical process (Whitworth \etal 1996), and in fact detailed calculations of ambipolar diffusion show that it is never closely realized since a dynamical collapse begins long before the singular state is reached (Basu and Mouschovias 1994; Mouschovias and Ciolek 1999; see Section~4.4).  Nevertheless, the simplicity and elegance of the SIS model and the ease with which it can be used to generate predictions have led to its wide use as a `standard model' for star formation, and it provides a useful reference model and limiting case.  More realistic models are expected to be intermediate between the two types of models that have been mentioned, which can be regarded as the limiting cases of fast and slow collapse.

\subsection{Spherical collapse} % 4.2

   In addition to the initial conditions, the dynamics of the collapse depends on the thermal behavior of the gas.  At low densities, the temperature is predicted to decrease with increasing gas density because of the increasing efficiency of atomic and molecular line cooling, while at higher densities the gas becomes thermally coupled to the dust, which then controls the temperature by its thermal emission; the temperature then begins to rise slowly with increasing density (Hayashi 1966; Larson 1973, 1985; Tohline 1982; Masunaga and Inutsuka 2000).  The net effect is that the temperature does not change much while the density increases by many orders of magnitude; the temperature in a collapsing core is predicted to remain in the range between about 6~K and 12~K as long as the core remains optically thin to the thermal emission from the dust, which is true for densities below about $10^{10}$~H$_2$~cm$^{-3}$.  Most collapse calculations have therefore assumed for simplicity that the early stages of collapse are isothermal with a constant temperature of 10~K.  This assumption is, however, somewhat crude and may not always be adequate; for example, the switch from molecular to dust cooling might have significant consequences for the details of fragmentation (Whitworth \etal 1998).

   Most calculations have adopted a fixed boundary for a collapsing cloud core, but other possibilities such as a constant boundary pressure have also been considered.  This does not change the qualitative nature of the collapse, and a universal result of calculations of isothermal collapse is that, regardless of the initial or boundary conditions, the collapse is always highly non-uniform and characterized by the runaway growth of a central density peak (Penston 1966; Bodenheimer and Sweigart 1968; Larson 1969).  This highly non-uniform nature of the collapse is of fundamental importance for the subsequent stages of evolution, and it occurs because the collapse of the outer layers is always slowed by an outward pressure gradient that develops when the interior pressure rises but the boundary pressure does not.  Even if the density and pressure are initially uniform, an outward pressure gradient is always created at the boundary when the collapse begins, and this gradient propagates inward as a rarefaction wave at the speed of sound (Bodenheimer and Sweigart 1968).  If gravity and pressure are initially approximately balanced, the rarefaction wave reaches the center before the collapse has progressed very far, and the density and pressure thereafter decrease monotonically outward.
  
   In the absence of a pressure gradient, the collapse of a uniform sphere of gas occurs in the free-fall time
\begin{equation}
                   t_{\rm ff} ~=~ (3\pi/32G\rho)^{1/2},            % (7)  
\end{equation}
defined as the time required to collapse to infinite density from a state of rest (Spitzer 1978).  In the presence of a finite outward pressure gradient, the collapse is somewhat decelerated from a free-fall, but the time required for each radial mass shell to collapse to the center is still approximately the free-fall time calculated from the average interior density of the shell.  As a result, the denser inner regions always collapse faster than the less dense outer regions, and the density distribution becomes increasingly centrally peaked (Larson 1973; Tohline 1982).

   Numerical calculations show that the density distribution in a collapsing iso\-thermal sphere approaches the asymptotic form $\rho \propto r^{-2}$ at progressively smaller radii as long as the isothermal approximation continues to hold (Larson 1969; Ogino \etal 1999).  A density distribution of this form, which is valid for an equilibrium isothermal sphere, is also approached in a dynamically collapsing sphere because the pressure gradient never becomes negligible near the center and prevents the central density peak from becoming too narrow at any stage; the peak width at any stage is always of the order of the Jeans length, and therefore is always proportional to $\rho^{-1/2}$.  The growth of the central density peak proceeds in a nearly self-similar way that is approximated by the similarity solution found by Penston (1969) and Larson (1969), which is valid asymptotically in the limit of high central densities and small radii where the initial and boundary conditions have been `forgotten'.  In the limit of infinite central density, this `Larson-Penston' (LP) similarity solution has a density distribution given by $\rho = 0.705\,c^2\!/Gr^2$ which is everywhere 4.43 times the density of the equilibrium SIS model, and it also has an asymptotically constant infall velocity equal to 3.28 times the sound speed.  Numerical collapse calculations show that the density distribution approaches the predicted $r^{-2}$ form after the central density has risen by several orders of magnitude, but the collapse velocity converges more slowly to the asymptotic value of the LP solution and reaches only about twice the sound speed before the isothermal approximation begins to break down.

   If the equation of state is not isothermal but can still be approximated by the polytropic form $P \propto \rho^{\gamma}$, qualitatively similar results are found that can be approximated by a similarity solution with the asymptotic form $\rho \propto r^{-2/(2-\gamma)}$ (Larson 1969; Ogino \etal 1999).  The effect of small departures from spherical symmetry on the collapse has been studied by Larson (1972a), Hanawa and Matsumoto (1999, 2000), and Lai (2000), with the conclusion that small departures from spherical symmetry may grow somewhat during isothermal collapse but probably not enough to alter greatly the qualitative results described above.  In particular, Larson (1972a) found that small departures from spherical symmetry tend to oscillate between prolate and oblate forms during the collapse, while a strongly prolate shape can become progressively more prolate and collapse to a thin spindle as predicted by Lin \etal (1965).  Hanawa and Matsumoto (1999, 2000) analyzed the stability of the LP solution to non-spherical perturbations and found a weak instability whereby prolate or oblate distortions grow slowly with increasing central density, but the amplitude of the distortion increases only as the 0.177 power of the central density, and this may not be enough to produce very large departures from spherical symmetry during the isothermal phase of collapse.

   A major implication of all of these results is that only a very small fraction of the mass of a collapsing cloud core first attains densities high enough to form a star, while most of the mass remains behind in an extended infalling envelope.  This is also true for the SIS model of Shu \etal (1987), in which the centrally peaked density distribution is assumed to develop quasi-statically; the essential difference is that in the SIS model the envelope is initially at rest, rather than falling inward at about twice the sound speed as in the dynamical collapse models.  In either case, a central stellar object or `protostar' with a very small initial mass is predicted to form at the center and to continue growing in mass by accretion from the surrounding envelope.  Thus the star that eventually forms at the center of a spherically collapsing cloud core is predicted to acquire almost all of its final mass by accretion from the infalling envelope; the accretion process will be discussed further in Section~5.  Mathematically, the formation of a star can be identified with the appearance of a singularity in the density distribution, and the accretion problem can be modeled by replacing the singularity with an accreting point mass and calculating its growth in mass as matter continues to fall into it.

\subsection{Collapse with rotation} % 4.3

  Most star-forming cloud cores are observed to be rotating (Goodman \etal 1993), as would be  expected because of the turbulence in molecular clouds (Burkert and Bodenheimer 2000), and this rotation must strongly influence the later stages of collapse if angular momentum is conserved.  The angular momentum of a typical prestellar cloud core is orders of magnitude more than can be contained in a single star, even if rotating at breakup speed, and this implies that some loss or redistribution of angular momentum is necessary if a star is to form; this is the classical `angular momentum problem' of star formation.  Early discussions of the angular momentum problem assumed that star-forming clouds acquire their angular momentum from galactic rotation, and this led to an angular momentum disparity of many orders of magnitude (Mestel and Spitzer 1956; Mestel 1965a,b; Spitzer 1968).  The observed rotation rates of prestellar cores are considerably smaller than is predicted on this basis, plausibly because magnetic fields have already carried away much of the initial angular momentum during the earlier low-density phases of cloud evolution when the fields remain strongly coupled to the gas (Mestel 1965a,b; Mouschovias 1977, 1991).  Nevertheless the observed angular momentum of prestellar cores is still about three orders of magnitude more than can be contained in a single star, and magnetic fields cannot dispose of all of this angular momentum because they are predicted to decouple from the gas and become dynamically unimportant during the later stages of the collapse (see Section~4.4).  In most cases, collapse with rotation probably results in the formation of a binary or multiple system of stars whose orbital motions can account for much of the initial angular momentum; the formation of such systems will be considered further in Section~6.  Here we consider first the more idealized case of collapse with axial symmetry.

   Several early efforts to calculate the collapse of an axisymmetric rotating cloud showed the formation of a ring with a central density minimum (Larson 1972a; Black and Bodenheimer 1976; Tohline 1980).  This type of configuration was hypothesized to be unstable to fragmentation into a binary or multiple system (Bodenheimer 1978), and this result was demonstrated numerically by Norman and Wilson (1978).  However, later calculations with finer spatial resolution did not yield a ring but rather a centrally condensed disk in which a central density singularity develops in an approximately self-similar way, qualitatively as in the non-rotating case (Norman \etal 1980; Narita \etal 1984).  Hayashi \etal (1982) and Toomre (1982) obtained an analytic solution for a singular isothermal disk that has the same distribution of angular momentum as a uniformly rotating cloud, and Hayashi \etal (1982) suggested that this singular disk approximates the end state of the collapse of a uniformly rotating isothermal cloud.  Detailed collapse calculations by Matsumoto \etal (1997) showed that such a singular disk is indeed approached, and that although the dynamics is complex in detail and involves recurrent shock formation, the evolution can be described in terms of oscillations around an asymptotic similarity solution that qualitatively resembles the LP solution with the addition of rotational flattening (Saigo and Hanawa 1998).  Calculations of non-isothermal collapse with rotation that assume various polytropic equations of state also show the approximately self-similar development of a central density singularity (Saigo \etal 2000).

   An important conclusion of this work is that rotation does not prevent the formation of a density singularity, which develops in qualitatively the same way as in the non-rotating case as long as the collapse remains isothermal.  As was emphasized by Narita \etal (1984), this behavior results essentially from the competition between pressure and gravity near the center, and centrifugal forces never become strong enough there to halt the increasing central condensation.  The main difference is that in the rotating case most of the mass ends up in a centrifugally supported disk around the central density singularity.  This disk is predicted eventually to become unstable or marginally stable according to the $Q$ criterion of Section~3.2, since $Q$ is predicted to become less than 1 in all cases and less than 0.4 in the most relevant cases (Hayashi \etal 1982; Larson 1984).  A possible result of marginal stability in such a disk is that spiral density fluctuations produce gravitational torques that transport angular momentum outward and drive an inflow onto the central object (Larson 1984), as occurs in the fully three-dimensional simulation of rotating collapse by Bate (1998).  Another possibility is that the disk eventually fragments to form a binary or multiple system of stars (Matsumoto and Hanawa 2003).  Narita \etal (1984) noted also that the outcome of rotating collapse is sensitive to departures from isothermality, and that a more realistic equation of state may sometimes lead to ring formation and hence fragmentation rather than to the development of a central density peak.

   When rotation is added to the SIS model of Shu \etal (1987), the result is that most of the envelope matter does not fall directly onto the central protostar but settles into a centrifugally supported disk around it (Terebey \etal 1984; Shu \etal 1987, 1993).  The growth in mass and radius of this circumstellar disk may involve a number of processes and stages, but the end result is again likely to be a gravitationally unstable or marginally stable disk (Stahler \etal 1994; Stahler 2000).  Again it is possible that spiral density fluctuations produce gravitational torques that drive inward mass transfer (Stahler 2000), but this requires that the growth of the density fluctuations must saturate before they become large enough to cause the disk to fragment, and it is not clear whether a steady state like this can be maintained.  Otherwise, the result may ultimately be fragmentation of the disk and the formation of a binary or multiple system.  Axisymmetric disk models may then be relevant only in situations where the disk never acquires enough mass to become self-gravitating.  The evolution of such protostellar disks will be discussed further in Section~5.2.

\subsection{Collapse with magnetic fields} % 4.4

   Because magnetic fields have been thought to be important in supporting molecular clouds against gravity (Section~2.3), much effort has been devoted to modeling the evolution of magnetically supported cloud cores that initially condense slowly by ambipolar diffusion (Nakano 1984; Shu \etal 1987; Mouschovias 1991; McKee \etal 1993; Mouschovias and Ciolek 1999).  The timescale for this ambipolar diffusion process is estimated to be of the order of $10^7$~years; since this is an order of magnitude longer than the free-fall time, the evolution is then expected to be quasi-static.  As the central part of such a magnetically supported cloud core slowly contracts, its self-gravity becomes increasingly important and it becomes increasingly flattened along the field lines.  Eventually gravity becomes strong enough to overwhelm magnetic support near the center and a runaway collapse ensues.  Detailed calculations of the evolution of a magnetized and rotating cloud core by Basu and Mouschovias (1994, 1995a,b) have shown that this dynamical collapse begins before the central density has increased by a very large factor, and typically before it has reached $10^5$~cm$^{-3}$; thus dynamical collapse is predicted to begin at densities not very different from those of observed prestellar cloud cores (Ciolek and Basu 2000).  The collapse accelerates as ambipolar diffusion continues to remove magnetic flux from the collapsing region, and it then proceeds qualitatively as in the non-magnetic case, with the runaway development of a central density singularity.  Because the collapsing region is flattened and retains some magnetic support, the collapse velocity is somewhat smaller than in the non-magnetic case, and it reaches a maximum value about equal to the sound speed.

   Nakano (1998) has questioned such ambipolar diffusion models, arguing that the observed prestellar cloud cores cannot be strongly magnetically supported because they would then not show the observed large enhancements in column density, and because it would then be difficult to account for their observed level of turbulence.  Nakano (1998) suggested that turbulence may play a more important role than magnetic forces in supporting cloud cores against gravity, and that dynamical collapse may be initiated more by the dissipation of turbulence than by ambipolar diffusion (see also Goodman \etal 1998; Myers and Lazarian 1998; Williams and Myers 2000).  Magnetic forces will then remain less important than gravity throughout the collapse, serving mainly to retard the collapse somewhat compared to the non-magnetic case (Indebetouw and Zweibel 2000; Heitsch \etal 2001).  In support of a relatively rapid collapse that is not much retarded by magnetic fields, Aikawa \etal (2003) find that the observed abundances of various molecules in star-forming cloud cores are most consistent with those predicted for rapidly collapsing cores, and that the agreement becomes worse if the collapse is greatly retarded by magnetic or other effects.

   Once a rapid dynamical collapse begins, a central density singularity develops in a nearly self-similar way even in the presence of a magnetic field.  Basu (1997, 1998) showed that the detailed results of Basu and Mouschovias (1994, 1995a,b) for the later stages of collapse of a rotating magnetized cloud core can be approximated by an asymptotic similarity solution that resembles the LP solution except for a flattened geometry and some continuing retardation by magnetic support; this solution again predicts an asymptotic density profile of the form $\rho \propto r^{-2}$, and it predicts an asymptotically constant collapse velocity of about twice the sound speed.  Detailed calculations by Tomisaka (1996a) and Nakamura \etal (1995, 1999) of the collapse of magnetized cloud cores formed by the fragmentation of a magnetized filament also show the formation of a flattened structure that develops a central density singularity in a nearly self-similar way; as in the case of rotating collapse, the dynamics is complex and involves the recurrent formation of shocks, but it can be described in terms of oscillations around an asymptotic similarity solution similar to that of Basu (1997).  Nakamura \etal (1999) suggested that this kind of self-similar collapse is universal and is approximated by essentially the same similarity solution as was found by Saigo and Hanawa (1998) for rotating collapse, with magnetic support here taking the place of centrifugal support.  Calculations of magnetic collapse that are continued through to the stage of accretion by a central point mass show that ambipolar diffusion is revived during the accretion phase in the weakly ionized inner part of the accreting envelope; this later phase of ambipolar diffusion may be responsible for removing most of the initial magnetic flux from the matter that goes into a forming star and hence for solving the `magnetic flux problem' of star formation (Ciolek and K\"onigl 1998; Contopoulos \etal 1998).

   Thus a magnetic field, like rotation, does not prevent the formation of a central density singularity when gravity gains the upper hand and causes a dynamical collapse to occur.  Even if the observed prestellar cloud cores were formed as a result of ambipolar diffusion, they are predicted to collapse dynamically almost from their observed state, and their later evolution is then only moderately retarded by the residual magnetic field.  The density distribution that results has the same asymptotic $r^{-2}$ form as in the equilibrium SIS model, but in the dynamically collapsing case the envelope is flattened and falling in at about the sound speed rather than spherical and static.  Some features of these results were incorporated in a generalization of the SIS model by Li and Shu (1996) that included flattening and magnetic support; these authors derived a solution for a singular isothermal magnetized disk that is similar to the Hayashi-Toomre disk, with magnetic support taking the place of centrifugal support.  The later evolution of the system may then involve accretion from this disk onto a central protostar, as will be discussed further in Section~5.

\subsection{Optically thick phases} % 4.5

   The results described above for spherical collapse and for collapse with rotation or magnetic fields show that thermal pressure never becomes negligible near the center, and that a central density peak always develops in qualitatively the same way, controlled by the competition between thermal pressure and gravity at the center.  The assumption of spherical symmetry might then provide an adequate approximation for the later stages of evolution of the central density peak when its optical depth becomes large and radiative cooling becomes unimportant, causing the central temperature to rise substantially.  The applicability of this assumption receives some support from the fully three-dimensional calculation of rotating collapse by Bate (1998); even though rotational flattening eventually becomes important and transient spiral features appear, gravitational torques transfer enough angular momentum outward to allow the continuing growth of a single central mass concentration that evolves in much the same way as in the spherical case.

   The calculations of Larson (1969, 1972b), Appenzeller and Tscharnuter (1975), Winkler and Newman (1980a,b), and Masunaga and Inutsuka (2000) for the later stages of spherical collapse have yielded similar results for the formation of a central stellar object or protostar, and somewhat similar results have been obtained also by Wuchterl and Tscharnuter (2003).  (Earlier calculations by Hayashi and Nakano (1965) and Bodenheimer (1968) had obtained results that were qualitatively similar but quantitatively different because they had assumed much higher initial densities.)  The central density peak becomes opaque to the thermal radiation from the dust grains when the central density reaches about $10^{-13}$~g~cm$^{-3}$ or $2 \times 10^{10}$ H$_2$~cm$^{-3}$, and the central temperature then begins to rise above its initial value of $\sim 10$~K.  A treatment of radiative transfer then becomes necessary, and various approximations have been used by the above authors, but they have yielded similar results for the transition from the initial isothermal phase to an adiabatic phase of evolution.  The gas becomes completely adiabatic at densities above $10^{-12}$~g~cm$^{-3}$, with a ratio of specific heats $\gamma \simeq 7/5$ that is appropriate for a gas consisting mostly of molecular hydrogen.  As the density continues to rise, pressure then increases faster than gravity and the collapse decelerates, essentially  coming to a halt when the central density reaches about $2 \times 10^{-10}$ g~cm$^{-3}$.  A central region that is nearly in hydrostatic equilibrium then forms, and it continues to grow in mass as matter falls into it through an accretion shock that develops at its surface.

   This first `hydrostatic core' has a mass of about 0.01~M$_\odot$ and a radius of several AU, and its properties are almost independent of the initial or boundary conditions because of the convergence toward self-similar behavior that occurs during the isothermal phase of collapse; this means that the properties of the central region eventually depend only on the thermal physics of the gas.  Since little if any fragmentation to smaller masses is likely to occur after an opaque hydrostatic core has formed, the mass of this first hydrostatic core is expected to be the minimum mass that can be achieved by fragmentation, and it is essentially the same as the `opacity limit' of about 0.007~M$_\odot$ first derived by Low and Lynden-Bell (1976).  The first hydrostatic core is however a transient feature and a second phase of collapse begins when the central temperature rises above 2000~K, causing hydrogen molecules to dissociate and reducing the value of $\gamma$ below the critical value of 4/3 required for stability.  This second phase of central collapse proceeds in qualitatively the same way as the earlier isothermal phase, and is again characterized by the runaway growth of a central density peak.  Rapid collapse continues until the hydrogen at the center is mostly ionized and the value of $\gamma$ has risen to a value near 5/3 that is characteristic of the ionized interior of a star.  The collapse is then permanently halted at the center and a second hydrostatic core forms, bounded again by an accretion shock into which matter continues to fall.  This second or `stellar' core initially has a very small mass of only about 0.001~M$_\odot$ and a radius of about 1~R$_\odot$, but it proceeds to grow rapidly in mass and also somewhat in radius, and as a result its initial properties are soon `forgotten' and have little effect on the later stages of evolution.  Within a very short time of only about 10 years, all of the mass of the first core has fallen into the second core or nascent protostar, but most of the initial collapsing mass still remains behind in an extended infalling envelope that continues to fall into the central protostar.

   The essential result of this work is the prediction that a star begins its life as a small `embryo' whose mass is less than $10^{-2}$ solar masses.  This embryo star or protostar continues to grow in mass as matter continues to fall onto it through the accretion shock at its surface.  When it first forms, the infalling matter outside the accretion shock is still optically thick and the shock is therefore adiabatic; as a result, the outer layers of the protostar are strongly heated and it expands rapidly.  After the material of the first hydrostatic core has been accreted, however, the opacity of the matter outside the accretion shock drops rapidly and radiation begins to escape freely from the shock.  Because of this radiative energy loss the protostar stops expanding, and it subsequently maintains an almost constant radius of about 4~R$_\odot$ during the remainder of the accretion process (Masunaga and Inutsuka 2000).  Eventually the accreting protostar becomes a normal pre-main-sequence star, which by then has acquired nearly all of its mass by accretion from the envelope.  Accretion is thus an essential part of the star formation process, and the accretion phase of evolution will be discussed further in Section~5.  The idea that stars might acquire most of their mass by accretion from the interstellar medium is actually an old one that predates nearly all of the work described above, having been suggested first by Hoyle and Lyttleton (1939).

   Most calculations of the adiabatic phase of collapse have assumed spherical symmetry, but the fully three-dimensional calculation by Bate (1998) of the collapse of a slowly rotating prestellar cloud core predicts the formation of a second hydrostatic core that has properties similar to those found in the spherical case.  When rotation is present, however, the remaining matter still has significant angular momentum and most of it will eventually settle into a disk around the central protostar; the later stages of evolution may then involve accretion from the disk (Yorke and Bodenheimer 1999).  Disk accretion may play the same role in early stellar evolution as spherical accretion if the outward transfer of angular momentum in disks is efficient enough to yield a similar accretion rate (Mercer-Smith \etal 1984).  The problem of disk accretion will be discussed further in Section~5.2.

\section{Accretion processes and early stellar evolution} % 5

   The calculations summarized above predict that a stellar object when first formed has a very small mass, and this implies that a star must acquire most of its final mass by accretion from a residual envelope.  In the spherical collapse calculations, the accreting envelope is initially falling inward at about twice the sound speed, whereas in the equilibrium SIS model it is initially static.  In the presence of rotation or magnetic fields, the innermost part of the envelope may become strongly flattened, and accretion may occur mostly from a disk; even more complex accretion geometries are possible in the more general situation where stars form in binary or multiple systems or clusters (see Section~6).  The geometry of the accretion process should not be very important for the internal evolution of an accreting protostar, however, since this depends mainly on how the protostellar mass increases with time.  The most important feature of the accretion process that needs to be understood is then the accretion rate as a function of time.  Because of the importance of accretion processes in star formation, much attention has been devoted to this subject, and it has been reviewed extensively by Hartmann (1998).

   As was emphasized by Stahler \etal (1980a,b, 1981), the study of protostellar evolution during the accretion phase can conveniently be separated into two problems, the first being that of modeling the accretion process and determining the accretion rate as a function of time, and the second being that of modeling the evolution of a protostar whose accretion rate is known from the solution of the first problem.  In the case of low-mass stars, the effect of the modest radiative output of the central protostar on the dynamics of the infalling envelope is small, and as a result, the accretion rate is almost independent of the evolution of the central protostar.  A bipolar outflow might eventually disperse part of the infalling envelope (Section~5.4), but such outflows are highly collimated and may not have much effect on the accretion process, especially if the envelope is flattened in a plane perpendicular to the outflow.  In the case of massive stars, however, the radiation from a luminous forming star can have much more important effects on the infalling envelope via heating, radiation pressure, and ionization; strong winds from massive forming stars may also help to disperse residual envelopes and limit or terminate the accretion process.  Therefore the formation of massive stars is usually considered separately from that of low-mass stars, and it will be discussed in Section~7.  Here we consider first the case where feedback effects are not important.

\subsection{Spherical accretion} % 5.1

   Spherical accretion without rotation or magnetic fields is a particularly simple problem to treat because thermal pressure is the only force counteracting gravity, and it is important mainly in the outer part of the envelope, which remains nearly isothermal.  A further simplification occurs because most of the mass in the inner envelope quickly falls into the central object and the gravitational field in this region then becomes that of a point mass.  An elegant similarity solution for accretion from a singular isothermal sphere onto a central point mass was derived by Shu (1977), who assumed that such a configuration begins to collapse at its center as soon as an accreting point mass has formed.  The infall region then grows with time in an `inside-out collapse' as a rarefaction wave propagates outward from the center at the speed of sound, converting the initial equilibrium configuration into a collapsing one as it goes.  The Shu (1977) similarity solution differs from the Larson-Penston solution described in Section~4.2 in that the LP solution approximates the formation of a central density singularity during isothermal collapse, while the Shu solution describes the post-singularity evolution of an assumed equilibrium SIS.  The accretion rate in the Shu solution is independent of time and depends only on the isothermal sound speed:
\begin{equation}
                        \dot M ~=~ 0.975\,c^3\!/G.                  % (8)
\end{equation}
For a temperature of 10~K this constant accretion rate is $1.53 \times 10^{-6}$ M$_\odot$~yr$^{-1}$; accretion at this rate would build a 1~M$_\odot$ star in $6.5 \times 10^5$~years.  Although the SIS model is not realistic in detail, as was seen in earlier, the inside-out collapse of the Shu solution qualitatively resembles the results of more realistic collapse calculations in that the latter also show the development of a growing region of near free-fall collapse; the Shu solution might then approximate part of the evolution of a real collapsing cloud core if its outer layers retain significant pressure support after a central protostar has formed.  The SIS model and its inside-out collapse with a simple constant accretion rate have provided the basis for a standard model of early stellar evolution that has been widely used and elaborated in the literature (Shu \etal 1987, 1993, 1999; Hartmann 1998).

   Although the LP solution was derived to approximate the pre-singularity evolution of a collapsing isothermal sphere, Hunter (1977) showed that it can be extended smoothly through the formation of a central singularity to a subsequent phase of accretion onto a central point mass.  This extended `Larson-Penston-Hunter' (LPH) solution predicts an accretion rate of 46.9$\,c^3\!/G$ which is much higher than that of the Shu solution because the infalling envelope of the LPH solution is denser by a factor of 4.43 than the equilibrium SIS and because it is not at rest but is falling inward at 3.28 times the sound speed.  Hunter (1977) demonstrated the existence of a family of similarity solutions with properties intermediate between those of the LPH and Shu solutions, and Whitworth and Summers (1985) showed that an infinite family of similarity solutions can be constructed for which the LPH solution and the Shu solution represent opposite limiting extremes.  To determine which case best approximates the behavior of a collapsing isothermal sphere, Hunter (1977) made test calculations that were continued into the accretion phase and found results that most resemble the LPH solution, although they approach it closely only in a small region near the center; the resulting post-singularity accretion rate is initially about 36$\,c^3\!/G$ and then declines rapidly with time.  Foster and Chevalier (1993) also made test isothermal collapse calculations that were continued throughout most of the accretion phase, and they found that the post-singularity accretion rate briefly approaches the LPH value but then declines strongly with time, eventually reaching values much smaller even than the Shu value as the infalling envelope becomes depleted.  In similar test calculations, Ogino \etal (1999) also found a  post-singularity accretion rate that is consistent with the LPH value, and they generalized these results to a non-isothermal equation of state, showing that the post-singularity accretion rate is a sensitive function of $\gamma$; for example if $\gamma > 1$, the accretion rate immediately after singularity formation is even higher than the LPH value and it then declines even more rapidly with time.

   The results of realistic collapse calculations that include an optically thick non-isothermal phase (Section~4.5) also approach the LP solution during the initial isothermal phase, but they never come very close to it before opacity intervenes and the isothermal assumption breaks down.  When the first hydrostatic core forms, the surrounding infalling envelope has a density about twice that of the equilibrium SIS and it is falling inward at about twice the sound speed.  These properties do not change much during the short lifetime of the first hydrostatic core, and therefore the accretion rate following the disappearance of the first core is considerably higher than the standard-model value of equation (8) but not as high as the LPH value, briefly reaching about 13$\,c^3\!/G$ and then decreasing with time as the envelope becomes depleted.  Typically, the initial burst of rapid accretion lasts less than $10^4$~years, while the time taken for half of the envelope to be accreted is about $10^5$~years; nearly all of the envelope is accreted within $10^6$~years.  By contrast, in the standard model the accretion rate remains constant indefinitely because the assumed SIS is unbounded and extends to indefinitely large radius and mass; other effects such as stellar outflows are in this case required to terminate the accretion process.

  As was seen in Sections 4.3 and~4.4, calculations of axisymmetric collapse with rotation or magnetic fields show that neither of these effects can prevent the runaway growth of a central singularity once a dynamical collapse has begun, and a central point mass still forms and begins to grow by accretion.  Basu (1998) showed that the approximate similarity solutions for magnetic isothermal collapse derived by Basu (1997) and Saigo and Hanawa (1998) can be extended to a post-singularity accretion phase and that they predict an initial accretion rate of about 25$\,c^3\!/G$, about half that of the LPH solution.  In detailed calculations of magnetic isothermal collapse carried into the accretion phase, Tomisaka (1996b) found that the post-singularity accretion rate is initially about 40$\,c^3\!/G$ and then declines strongly with time.  For the analogous problem of rotating isothermal collapse, Matsumoto \etal (1997) estimated a post-singularity accretion rate of about \hbox{(13--20)}$\,c^3\!/G$, while Nakamura (2000) found a post-singularity accretion rate that exceeds 30$\,c^3\!/G$ but then declines rapidly with time.  Thus all of the existing calculations of dynamical collapse, including those that incorporate rotation and magnetic fields, agree in predicting a post-singularity accretion rate that is initially much higher than the standard-model value of equation (8) and that subsequently declines strongly with time.  In simulations of star formation in clusters, the accretion rate is also found to be highly time-variable, typically declining rapidly after an initial peak (Klessen 2001a).

   Throughout most of the accretion phase, the central protostar is predicted to remain heavily obscured by dust in the infalling envelope; in the spherical case the envelope remains optically thick until almost all of its mass  has been accreted or dispersed.  Therefore it is difficult to observe most of the accretion process or estimate protostellar accretion rates from observations, but indirect evidence from protostellar luminosities and outflow rates suggests that the accretion rate is indeed very high during the earliest and most heavily obscured phases of protostellar evolution and declines strongly with time during the later stages (Hartmann 1998; Andr\'e \etal 1999, 2000).  Many visible T~Tauri stars show spectroscopic evidence for continuing infall, but the measured accretion rates for these stars are several orders of magnitude smaller than those that must have characterized the main accretion phase, typically being only about $10^{-7}$ or $10^{-8}$ M$_\odot$~yr$^{-1}$ (Hartmann 1998).  This low accretion rate no longer represents a significant rate of mass gain for these stars, so that by the time a visible star is seen, it has accreted virtually all of its final mass and the star formation process is essentially finished.

\subsection{Disk accretion} % 5.2

   In general, accretion will not be spherical because rotational flattening and disk formation eventually become important; even a slowly rotating prestellar cloud core has too much angular momentum for all of its mass to fall directly into a forming star, and some of this matter will then almost certainly form a circumstellar disk.  Even if most of the angular momentum of a prestellar cloud core goes into the orbital motions of the stars in a binary or multiple system (Section~6), the material accreted by each star will still have enough `spin' angular momentum to form a circumstellar disk, as is illustrated by the simulations of Bate (2000).  Indeed, theory, simulations, and observations all suggest that circumstellar disks are a very frequent if not ubiquitous feature of star formation (Hartmann 1998).  If most of the mass acquired by a forming star first settles into a centrifugally supported disk, it must then be transported inward through the disk to be accreted by the star, and this requires that its angular momentum must somehow be removed or transported outward through the disk.  Such outward transport of angular momentum can occur if the disk is viscous or if some mechanism creates an effective viscosity in the disk, as has been assumed in most of the models of accretion disks that have been studied in many contexts, including star formation, following Shakura and Sunyaev (1973) and Lynden-Bell and Pringle (1974).  The role and the possible mechanisms of disk accretion in star formation have been reviewed by Hartmann (1998).

   The central problem in the theory of accretion disks is to understand the mechanism(s) responsible for the assumed outward transport of angular momentum.  It has long been clear that molecular viscosity is many orders of magnitude too small to be important, and therefore macroscopic transport processes must operate if disk accretion is to occur.  Despite decades of work on the problem, no mechanism has yet been identified that is clearly capable of providing the desired outward transport of angular momentum, but some of the possibilities have been reviewed by Larson (1989), Adams and Lin (1993), Papaloizou and Lin (1995), Stahler (2000), and Stone \etal (2000).  The forces mainly responsible for transporting angular momentum can be purely hydrodynamic, gravitational, or magnetic, and in each case both small-scale random processes and large-scale ordered phenomena can play a role.  The initial suggestion that hydrodynamic turbulence might replace molecular viscosity now seems unlikely to be correct because centrifugally supported disks are stable against the spontaneous development of turbulence, and because such turbulence tends in any case to transport angular momentum {\it inward}, not outward (Stone \etal 2000; Quataert and Chiang 2000).  Another much-studied possibility is that weak gravitational instabilities in a marginally stable disk may generate transient spiral density fluctuations whose gravitational torques transport angular momentum outward (Larson 1984; Adams and Lin 1993; Bodenheimer 1995; Laughlin and R\'o\.zyczka 1996; Nomura and Mineshige 2000; Stahler 2000; Stone \etal 2000; Gammie 2001).  This effect can be significant if the mass of a disk is sufficiently large for self-gravity to be important, i.e.\ a few tenths of the mass of the central star; although this might be true for some of the youngest protostellar disks that are still heavily obscured, most observed protostellar disks have masses that are at least an order of magnitude smaller than this and therefore too small for self-gravitational effects to be important (Hartmann 1998; Beckwith 1999; Mundy \etal 2000).

   In recent years much attention has focused on the magnetorotational or `Balbus-Hawley' instability of magnetized disks, which can be much more effective than purely hydrodynamic turbulence as a transport mechanism if the degree of ionization is sufficient and magnetic coupling is important (Balbus and Hawley 1998; Stone \etal 2000).  The outermost part of a protostellar disk may be kept sufficiently ionized by cosmic rays, and a central region may be ionized by radiation from the central star, but most of the mass of a typical protostellar disk lies in an intermediate `dead zone' where ionization is negligible and this mechanism cannot operate except possibly in a surface layer (Gammie 1996).  Even if a larger region around the central star could be kept sufficiently ionized by self-sustaining MHD turbulence, most of the mass in a protostellar disk would still be in a dead zone (Fromang \etal 2002).  In any case this mechanism produces accretion rates that are only marginally sufficient to be important for star formation (Stahler 2000), so it too does not clearly provide the desired transport of angular momentum.  Another possibility is that a large-scale magnetic field threading a protostellar disk removes angular momentum via a centrifugally driven `disk wind' (Ouyed and Pudritz 1999; K\"onigl and Pudritz 2000), but the assumptions required for this mechanism to work are difficult to justify (Hartmann 1998), and no fully self-consistent model of this type has yet been constructed.  Centrifugally driven winds might be expected to be most important in the inner, most highly ionized part of a protostellar disk, and a model in which such a wind is driven entirely from the inner edge of a disk has been developed by Shu \etal (2000) to explain protostellar outflows.  However, this kind of wind cannot play a major role in removing angular momentum from a disk because the material at the inner edge of the disk must already have lost most of its angular momentum in some other way.

   It is therefore not clear that any mechanism intrinsic to a protostellar disk can drive significant accretion from the disk onto the central star, although some of the mechanisms mentioned above could play a role in some regions or circumstances.  Since the most direct evidence that we have for protostellar disks is for residual disks of low mass around stars whose formation has essentially been completed (Hartmann 1998; Beckwith 1999; Mundy \etal 2000), it is possible that disks like those postulated in standard models play a less central role in the star formation process than has been assumed, and that disks are more of a byproduct of complex and chaotic star formation processes than an essential feature (see Section~6).  Another possibility is that external perturbations are responsible for driving accretion from disks.  In a forming binary system, for example, the tidal effect of a companion star can create spiral disturbances in a circumstellar disk which propagate inward as acoustic waves; such tidally generated waves carry negative angular momentum and thus tend to reduce the angular momentum of the region in which they propagate (Spruit 1987, 1991; Larson 1989, 1990a,b; Lin and Papaloizou 1993; Lubow and Artymowicz 2000; Stone \etal 2000; Blondin 2000).  A desirable feature of waves as a possible transport mechanism is that any wave with a trailing spiral pattern always transports angular momentum outward, regardless of the nature or direction of propagation of the wave (Larson 1989).  Many numerical simulations have shown the formation of trailing spiral wave patterns in tidally perturbed disks, typically with a two-armed symmetry reflecting the symmetry of the tidal distortion (Sawada \etal 1987; R\'o\.zyczka and Spruit 1993; Savonije \etal 1994; Bate 2000; Makita \etal 2000; Blondin 2000).  These tidally generated waves often develop into shocks, and the associated dissipation can then permanently reduce the energy and angular momentum of the disk and drive an inflow (Shu 1976; Spruit \etal 1987; Larson 1989, 1990b; Spruit 1991; Boffin 2001).

   The importance of tidal waves in disks depends on many details of wave propagation and dissipation that are not currently well understood (Lubow and Artymowicz 2000; Bate \etal 2002c).  If such waves can propagate far enough before being completely damped out, they can potentially drive an inflow through the entire region in which they propagate.  Waves can also be generated within a disk by objects that form in the disk itself, such as massive planets or smaller companion stars; such objects can form by local gravitational instabilities if the disk is sufficiently massive and if radiative cooling is important (Gammie 2001; Boss 2001, 2002).  Even an object with a mass as small as that of Jupiter can have a significant influence on the evolution of a disk by tidally extracting angular momentum from the inner part and transferring it to the outer part (Goldreich and Tremaine 1982; Lin and Papaloizou 1993; Goodman and Rafikov 2001), potentially driving disk evolution on a timescale of $10^6$ years (Larson 1989).  Since the tidal torque between such an object and a disk is proportional to the square of the object's mass, objects more massive than Jupiter will produce much stronger effects, so that planets more massive than Jupiter can have major effects on the evolution of disks.  Goodman and Rafikov (2001) have suggested that the combined effect of many small planets could also provide an effective viscosity for a protostellar disk.

   It is also possible that several of the mechanisms described above could interact in very complicated ways, and that hydrodynamic, gravitational, and magnetic effects could all play some role in disk evolution.  The evolution of protostellar disks might then be a complex and chaotic process that is difficult to describe with simple models.  Even relatively simple physics can quickly lead to chaotic behavior if the simplifying assumptions usually adopted in theoretical models are relaxed, as will be discussed further in Sections 6 and~7.

\subsection{Early stellar evolution} % 5.3

   For a star with a final mass of the order of one solar mass, the initial protostellar mass is less than 1\% of its final mass, so the properties and evolution of the resulting star depend almost entirely on the 99\% or more of the mass that is acquired by accretion.  The collapse calculations described above yield a much lower initial specific entropy for a protostar than is typical for stars, owing to the strong radiative cooling that occurs at the center during the isothermal collapse, but the matter that is accreted subsequently by the protostar is heated to a much higher specific entropy by its passage through the accretion shock.  This yields a configuration that for a time has a central temperature minimum, but the small low-entropy region at the center is of minor importance for the structure of the protostar and it eventually disappears because of radiative heating from the surroundings before the star begins hydrogen burning, playing no important role in its later evolution.  Therefore the approximations used by Stahler \etal (1980a,b, 1981) and many subsequent authors to study protostellar evolution, bypassing the collapse phase and starting with simple models for the initial accreting protostar, introduce no serious errors and can be used to calculate the early stages of stellar evolution with reasonable accuracy, given a knowledge only of the accretion rate as a function of time.

   When an accreting protostar reaches a mass of about 0.2~M$_\odot$, deuterium burning begins at the location of the temperature maximum and becomes a significant heat source that keeps the protostar from contracting as its mass continues to increase.  For a range of assumptions about the history of the accretion process, the final radius when accretion ceases to be important is typically predicted to be about 4~R$_\odot$ for a star of mass 1~M$_\odot$ (Stahler \etal 1980a,b; Mercer-Smith \etal 1984; Stahler 1988; Hartmann \etal 1997; Masunaga and Inutsuka 2000).  The earlier calculations of Larson (1969, 1972b) and Winkler and Newman (1980a,b) had yielded a smaller radius of about 2~R$_\odot$ because they did not include deuterium burning, but apart from this difference in the final protostellar radius, all of these calculations yield a qualitatively similar picture for the early stages of stellar evolution.  They all show that after accretion has ceased to be important, a newly formed star of low mass has a structure similar to that of a conventional pre-main-sequence star with a convective envelope that lies on or near the lower part of the `Hayashi track' in the Hertzsprung-Russell diagram, a type of structure first studied by Hayashi \etal (1962).

   The earliest stages of stellar evolution have been reviewed by Hayashi (1966), Bodenheimer (1972), Lada (1991), Stahler and Walter (1993), Stahler (1994), Hartmann (1998), and Palla (1999, 2001, 2002), and will only be briefly summarized here.  Throughout almost the entire accretion phase, a protostar remains obscured by dust in the infalling envelope and the system is observable only at infrared wavelengths.  Initially the warmer inner part of the envelope is opaque even at near-infrared wavelengths, and only the cooler outer part radiates freely into space, causing the object to be observed as a far-infrared or submillimeter source.  As the envelope becomes depleted of matter, the optically thick region shrinks in size and the spectrum of emitted radiation shifts toward shorter wavelengths, until eventually the central star begins to shine through and a composite spectrum with both visible and infrared components is seen.  Observers have developed a simple classification scheme for this sequence of stages of evolution consisting of Classes 0, I, II, and~III, which designate objects whose dominant emission is at submillimeter, far infrared, near infrared, and visible wavelengths respectively.  The correspondence with the accretion history discussed in Section~5.1 is roughly that Class~0 corresponds to an early phase of rapid accretion lasting a few times $10^4$ years, Class~I to the main accretion phase lasting a few times $10^5$ years, Class~II to the appearance of a classical T~Tauri star with significant circumstellar dust, a stage lasting up to $10^6$ years, and Class~III to a `weak-line T~Tauri star' that no longer has any significant circumstellar material.  The available evidence suggests that the accretion rate declines from a rate much higher than the standard-model value for the Class~0 objects to a rate much lower than that for the T~Tauri stars, which are no longer gaining mass at a significant rate (Hartmann 1998; Hartmann \etal 1998; Andr\'e \etal 1999, 2000).

   The models described above predict that newly formed stars of low or moderate mass have similar radii when accretion becomes unimportant, the final protostellar radius increasing only modestly with mass.  Therefore newly formed stars should first appear along a locus of nearly constant radius in the Hertzsprung-Russell diagram.  Stahler (1983) called this locus the `birthline', denoting the fact that it is the locus along which stars are predicted to make their first visible appearance after emerging from their birth clouds.  (A similar locus had earlier been derived by Larson (1972b) with radii that were about a factor of 2 smaller because of the neglect of deuterium burning.)  After appearing on the birthline, stars contract for a few tens of millions of years until they become hot enough at their centers to burn hydrogen and settle into the long-lived `main sequence' phase of evolution.  Most observed young stars have radii smaller than that of the birthline, as expected, and their basic properties appear to be generally well accounted for by conventional models of early stellar evolution. These models have been used to derive the distribution in mass and age of the newly formed stars in different star forming regions, and an important result, already alluded to in Section~2, is that the age span of the newly formed stars associated with each star-forming cloud is small, typically only a few million years, showing that star-forming clouds are short-lived and that star formation is a rapid process.

   While the models described above may account satisfactorily for the most basic features of early stellar evolution, the observations have revealed some unexpected properties of young stars that were not predicted by any of the models.  One is that most newly formed stars or protostars show evidence for outflows which may be very energetic and may even dominate the observed properties of these young objects, being much more conspicuous than any signs of infall or accretion (Edwards \etal 1993; Eisl\"offel \etal 2000).  Jet-like bipolar outflows may provide the first sign that an accreting protostar has formed in a prestellar cloud core, since they appear already very early during the first $10^4$ years of evolution when a protostar is still heavily obscured (Reipurth 1991; Fukui \etal 1993).  The energy source for these jets is believed to be the gravitational energy of the matter accreted by a central protostar, and their collimation is believed to be caused by a helical magnetic field that is coupled either to the inner part of a circumstellar disk or to the central protostar.  It seems almost certain that both rotation and magnetic fields are involved in the origin of the bipolar outflows, although the various models that have been proposed differ considerably in their details (K\"onigl and Ruden 1993: K\"onigl and Pudritz 2000; Shu \etal 2000; Tomisaka 2002).

   Another feature of early stellar evolution that was not predicted is that newly formed stars often show strong variability in luminosity that may reflect, at least in part, variability of the accretion process (Hartmann \etal 1993).  The jet activity is also variable, and the jets appear to be emitted in spurts; it is possible, but not presently established, that their production is associated with flareups in stellar luminosity, the most extreme examples of which are the `FU~Orionis' outbursts (Herbig 1977; Hartmann and Kenyon 1996; Herbig \etal 2003).  The observations thus show clearly that even though we may have in hand the basic elements of a theoretical understanding of early stellar evolution, star formation is a more complex and chaotic process than in any of the models that have been discussed so far.  One reason for the greater complexity is almost certainly that most stars do not form in isolation but in binary or multiple systems where interactions can play an important role, as will be discussed further in Section~6.

\subsection{Accretion and stellar masses} % 5.4

   If stars acquire most of their mass by accretion, then understanding the origin of stellar masses requires understanding in more detail the history of the accretion process itself and what terminates it.  In the spherical collapse calculations described above, the collapsing region is bounded in size and mass, and all of its mass eventually falls into the central protostar.  Most of this mass is still expected to end up in a central star or binary system even if angular momentum is important and a disk forms, because if too much mass goes into a circumstellar disk, the effects of self-gravity in the disk will become important and either redistribute angular momentum until most of the mass is in a central star (Larson 1984; Bodenheimer 1995; Stahler 2000) or cause the disk to fragment into a binary or multiple system.  If a cloud core is partly supported by a magnetic field, its outer layers may retain some magnetic support even after the inner part has collapsed, and this may reduce the efficiency with which matter condenses into a central star (Mouschovias 1990).  However, in models in which the parameters are chosen to achieve the best consistency with observations (Ciolek and Basu 2000, 2001), magnetic fields play a less dominant role than in some earlier models and may be less important in reducing the efficiency of star formation, although quantitative predictions have not yet been made.

   Real prestellar cloud cores, in any case, are not sharply bounded but merge smoothly into their surroundings, and the part of a cloud core that will eventually go into a star may also not be very well defined.  Some studies of the radial density profiles of prestellar cloud cores have suggested that the density falls off sufficiently rapidly at the edge that the size and mass of a core can be reasonably well defined (Motte \etal 1998; Andr\'e \etal 2000, 2003; Ward-Thompson 2002).  Motte \etal (1998) and Motte and Andr\'e (2001a) found that when core masses are measured in this way, their distribution of masses resembles the stellar initial mass function, and they suggested on this basis that these cores represent direct stellar progenitors that collapse into stars with similar masses.  Other authors, using different techniques for measuring core masses, have found qualitatively similar results but core masses that are systematically larger by a factor of 2 or 3 (Johnstone \etal 2000, 2001).  Earlier studies of the `ammonia cores' in many star-forming clouds had also suggested that these cores could be identified as direct stellar precursors with masses similar to those of the stars that form in them (Myers 1985; Lada \etal 1993), although in this case it is possible that the ammonia molecule does not probe all of the mass present.  In all, it appears that prestellar cloud cores can be identified in which a significant fraction of the mass ends up in stars, although sizeable quantitative uncertainties still exist.  To the extent that stellar masses derive directly from the masses of these prestellar cloud cores, the cloud fragmentation processes discussed in Section~3 will contribute to determining stellar masses via fragmentation scales such as the Jeans mass; the fact that the Jeans mass in the densest regions is typically of the order of one solar mass may then help to explain why a typical stellar mass is also of this order, or slightly less.

   The `standard model' based on assuming an equilibrium SIS for the initial state does not directly provide any basis for predicting stellar masses, since the SIS is a scale-free configuration and the resulting accretion rate remains constant indefinitely.  In this case it is necessary to assume that other physical effects eventually terminate the accretion process, and it has usually been assumed that this is done by protostellar outflows, possibly triggered by the onset of deuterium burning in an accreting protostar (Shu \etal 1987, 1993, 1999).  Models that invoke outflows to terminate the accretion process and thereby account for the origin of stellar masses have been developed by a number of authors including Nakano \etal (1995) and Adams and Fatuzzo (1996), and have been reviewed by Meyer \etal (2000).  These models tend to contain many assumptions and parameters, making the role of outflows in determining stellar masses uncertain.  Outflows could plausibly help to disperse protostellar envelopes and thereby limit the amount of matter accreted by a protostar, and there is indeed evidence for dispersal of cloud material by outflows (e.g., Arce and Goodman 2002); however, the fact that outflows tend to be strongly collimated suggests that most of the matter in protostellar envelopes may not be much affected by them.  To the extent that outflows do play a role, they could be one of several factors, along with rotation and magnetic fields, that tend to reduce the efficiency of star formation.  If stellar masses are determined by a number of factors that vary independently and randomly, the resulting stellar mass spectrum may approach a lognormal form (Zinnecker 1984; Adams and Fatuzzo 1996), and such a form may provide a reasonable fit to the observed mass function of low-mass stars (Miller and Scalo 1979; Scalo 1986, 1998; Meyer \etal 2000; Kroupa 2001, 2002).  If a typical stellar mass is a few times smaller than a typical prestellar core mass, as some of the observations discussed above suggest, then variable efficiency factors depending on a number of different mechanisms could shift and broaden the stellar mass spectrum and help to account for its roughly lognormal form at low masses.  In this case, molecular cloud properties may determine the order of magnitude of stellar masses via mass scales such as the Jeans mass, while other effects including rotation, magnetic fields, and outflows could contribute in varying degrees to determining the detailed form of the IMF.

\section{Formation of binary systems} % 6

\subsection{The angular momentum problem} % 6.1

   As has long been recognized, the amount of angular momentum in a typical star-forming cloud core is several orders of magnitude too large to be contained in a single star, even when rotating at breakup speed; this is the classical `angular momentum problem' of star formation (Mestel and Spitzer 1956; Mestel 1965a,b; Spitzer 1968; Bodenheimer 1978, 1995.)  The angular momentum problem remains despite the fact that star-forming cloud cores rotate more slowly than would be expected if they had formed from low-density gas with conservation of angular momentum (Goodman \etal 1993); this slow rotation can plausibly be explained as a consequence of magnetic braking acting during the early low-density phases of cloud evolution when magnetic fields are still strongly coupled to the gas (Mestel 1965a,b; Mouschovias 1977, 1991).  Magnetic fields are, however, predicted to decouple from the gas by ambipolar diffusion long before stellar densities are reached, and angular momentum is then expected to be approximately conserved during the later stages of collapse (Basu and Mouschovias 1995a,b; Basu 1997; Mouschovias and Ciolek 1999).  Observations of collapsing cloud cores confirm that angular momentum is approximately conserved on scales smaller than a few hundredths of a parsec (Ohashi \etal 1997; Ohashi 1999; Myers \etal 2000; Belloche et al 2002).  Therefore an important part of the angular momentum problem remains unsolved, since the typical observed angular momentum of prestellar cloud cores is still three orders of magnitude larger than the maximum amount that can be contained in a single star (Bodenheimer 1995).

   In principle, there are two ways to dispose of this excess angular momentum: (1) it could be transported to outlying diffuse material, for example by viscous transport processes in a protostellar accretion disk, thus allowing most of the mass to be accreted by a central star, or (2) much of the initial angular momentum could go into the orbital motions of the stars in a binary or multiple system.  As was noted in Section~5.2 and discussed by Larson (2002), transport processes in disks do not clearly offer a solution to the angular momentum problem because no adequate transport mechanism has yet been identified, and because even if such a mechanism could be identified, a residual disk would have to expand to a very large radius to absorb all of the angular momentum, and this would make the accretion time longer than the inferred ages of young stars.  The alternative possibility, which has also long been recognized, is that the collapse of a prestellar cloud core typically produces a binary or multiple system and that much of the initial angular momentum goes into the orbital motions of the stars in such a system (Mestel and Spitzer 1956; Larson 1972a; Mouschovias 1977; Bodenheimer 1978).  Observations show that the great majority of stars do indeed form in binary or multiple systems (Heintz 1969; Abt 1983; Duquennoy and Mayor 1991; Mathieu 1994; Looney \etal 2000; Zinnecker and Mathieu 2001); since the angular momentum of a typical cloud core is comparable to that of a wide binary (Bodenheimer \etal 1993; Simon \etal 1995; Bodenheimer 1995), it is plausible that at that at least the wider binaries can be formed directly by the fragmentation of rotating cloud cores.  Even the minority of stars that are observed to be single can be accounted for if stars typically form in unstable triple systems that decay into a binary and a single star (Larson 1972a; Reipurth 2000); this simple scenario would be consistent with the observed fact that about two-thirds of all stars are in binary systems while about one-third are single.

\subsection{Formation of binary and multiple systems} % 6.2

   Much theoretical and numerical work has suggested that the formation of binary or multiple systems is the usual result of collapse with rotation, while the formation of single stars occurs only in special cases (Larson 1978; Boss 1990, 1993a,b; Burkert \etal 1997; Bonnell 1999; Bodenheimer \etal 1993, 2000; Zinnecker and Mathieu 2001; Sigalotti and Klapp 2001; Bate \etal 2002a,b, 2003).  In many cases a cloud core with significant rotation may fragment during the isothermal phase of collapse and directly form a binary or multiple system.  A systematic study of this problem by Tsuribe and Inutsuka (1999a,b) concludes that the occurrence of fragmentation during isothermal collapse depends primarily on the initial ratio of thermal to gravitational energy, which is related to the number of Jeans masses present, and only secondarily on the rotation rate; fragmentation is predicted if this ratio is less than about 0.5.  If collapse continues into an adiabatic phase with an opaque core and a flattened rotating configuration forms, such as a disk or bar, most of the initial mass will then settle into this flattened configuration and it will become unstable to fragmentation unless there is efficient outward transport of angular momentum.  An extensive exploration of this problem by Matsumoto and Hanawa (2003) shows that, in a wide range of cases, the final outcome is the formation of a binary or multiple system, which can occur via the formation and breakup of a ring or bar configuration as well as by the fragmentation of a disk.

   If a central star of significant mass forms with a residual disk of only modest mass, the final outcome is less clear, but recent simulations that include a detailed treatment of radiative cooling show that even in a disk of modest mass, fragmentation can occur and lead to the formation of a smaller companion object such as a massive planet or small star (Boss 2001, 2002; Rice \etal 2003).  If collapse begins with initial configurations that are far from axisymmetric, for example if the initial configuration is filamentary, or if a collapsing cloud core is perturbed by interactions with other cloud cores in a forming group or cluster, the formation of a binary or multiple system is even more likely to occur (Bonnell 1999; Bodenheimer \etal 2000; Whitworth 2001).  In the most detailed and realistic simulation of cloud collapse and fragmentation yet performed, which models the formation of a small cluster of stars in a turbulent and filamentary collapsing cloud (Bate \etal 2002a,b, 2003), most of the stars form in several subgroups, and within these groups numerous binary and multiple systems are formed; in fact, the dynamics of these groups is highly complex and chaotic, and binary and multiple systems are continually formed and disrupted.  Circumstellar disks are also continually formed and disrupted, and although they appear rather ubiquitously, they may often have only a transient existence before being either fragmented or disrupted by interactions.

   Observed binary systems show a large dispersion in their properties, and they have no strongly preferred values for parameters such as separation and eccentricity; this observed broad dispersion itself suggests a very dynamic and chaotic formation process (Larson 2001).  Although the wider binaries might plausibly be formed directly by the fragmentation of rotating cloud cores, the large spread in binary separations toward smaller values implies that additional processes must be involved in the formation of the closer binaries; these processes must have a stochastic element to account for the large dispersion in separations, and also a dissipative element to reduce the average energy and angular momentum of typical forming binaries (Larson 1997, 2001; Heacox 1998, 2000).  Purely stellar-dynamical interactions in groups and multiple systems can account for some of the observed spread in binary properties, but they cannot account for the closest observed systems (Sterzik and Durisen 1998, 1999; Kroupa and Burkert 2001).  The detailed simulation of cloud collapse and fragmentation by Bate \etal (2002a,b, 2003) produces many close binary systems with separations smaller than would be predicted by simple arguments based on conservation of angular momentum, and Bate \etal (2002b) conclude that three distinct mechanisms are involved in the formation of these close binaries: (1) the continuing accretion of gas by a forming binary, which can shrink its orbit by a large factor if the accreted gas has a smaller specific angular momentum than the binary, as is usually the case; (2) the loss of angular momentum from a forming binary due to tidal interaction with circumbinary gas, for example with gas in a circumbinary disk; and (3) dynamical interactions in forming multiple systems that tend to extract angular momentum from the closer binary pairs.  Clearly, many complex processes may be involved in the formation of binary and multiple systems, and large-scale simulations are needed to address this problem in a systematic way and make statistical predictions that can be compared with observations.

\subsection{The role of tidal interactions} % 6.3

   If most stars form in binary or multiple systems, gravitational interactions between the orbiting protostars and residual gas will play an important role in redistributing angular momentum in the system.  Tidal interactions, for example, can redistribute angular momentum in several different ways: (1) angular momentum can be extracted from a circumstellar disk by tidal interaction with a companion star in a binary or multiple system; (2) an object that forms in a circumstellar disk, such as a giant planet or a small companion star, can tidally extract angular momentum from the inner part of the disk and transfer it to the outer part; or (3) a forming binary system can lose orbital angular momentum by interaction with surrounding material, for example with matter in a circumbinary disk.

   If each accreting protostar in a forming binary system has its own circumstellar disk, as might be expected and as occurs in numerical simulations (Bate and Bonnell 1997; Bate 2000), the tidal perturbing effect of the companion star on each disk produces a gravitational torque that extracts angular momentum from the disk and transfers it to the binary orbit (Lubow and Artymowicz 2000; Bate 2000; Nelson 2000).  The transfer of angular momentum from the disk to the orbit can drive continuing accretion from each disk onto its central protostar, and this may be an important mechanism for driving accretion from disks onto forming stars (Larson 2002).  A tidal perturbation generates a two-armed spiral disturbance in a disk that propagates inward as an acoustic wave, and such a wave can transport angular momentum outward in the disk, possibly in conjunction with other mechanisms as was discussed in Section~5.2.  In a binary system, tidal interactions will tend to be self-regulating because if too little angular momentum is removed from a circumstellar disk, the disk will expand toward the companion as it gains matter, increasing the strength of the interaction; conversely, if too much angular momentum is removed, the disk will shrink and the tidal effect will be weakened.  Numerical simulations show that the circumstellar disks in a forming binary system settle into a quasi-steady state in which angular momentum is continually transferred from the disks to the binary orbit (Bate and Bonnell 1997; Bate 2000), although the internal transport of angular momentum within the disks is not accurately modeled in these simulations and may result partly from an artificial viscosity.  Tidal transport mechanisms may be of quite general importance in astrophysics, and may drive accretion flows in many types of binary systems containing disks (Matsuda \etal 2000; Menou 2000; Blondin 2000; Boffin 2001).

   If the orbit of a forming binary system is eccentric, as is true for most binaries, the tidal effect will be time dependent and will produce strong disturbances at each periastron passage (Bonnell and Bastien 1992), possibly causing episodes of rapid accretion onto one or both stars.  For stars that form in multiple systems or clusters, protostellar interactions can be even more violent and chaotic and can cause circumstellar disks to be severely disturbed or even disrupted (Heller 1995; Bate \etal 2003).  In such situations, protostellar accretion rates will vary strongly with time, regardless of the detailed mechanisms involved, and protostars may gain much of their mass in discrete events or episodes.  Some of the evidence suggests that protostellar accretion rates are indeed highly variable and that much of the accretion may occur in bursts.  The observed luminosities of many protostars are lower than is predicted for models with steady accretion, and this can be understood if most of the accretion occurs in short bursts (Kenyon and Hartmann 1995; Calvet \etal 2000).  The jet-like Herbig-Haro outflows, which are believed to be powered by rapid accretion onto forming stars at an early stage of evolution, are clearly episodic or pulsed, and this suggests that the accretion process is itself episodic (Reipurth 2000, 2001).  It is noteworthy that the jet sources are frequently found to have close stellar companions: at least 85 percent of the jet sources are members of binary or triple systems, which is the highest binary frequency yet found among young stellar objects (Reipurth 2000, 2001).  This fact strongly suggests that there is a causal connection between the presence of a close companion and the launching of a jet, as would be expected if tidal interactions were responsible for the episodes of rapid accretion that generate the jets (Reipurth 2001; Larson 2001, 2002).

   The same bursts of rapid accretion that produce the jets might also be responsible for the `FU~Orionis' outbursts that are observed in some newly formed stars; these outbursts are also thought to be caused by episodes of rapid accretion (Dopita 1978; Reipurth 1989, 2000; Hartmann and Kenyon 1996).  The possibility that the FU~Orionis phenomenon is caused by tidal interactions in highly eccentric binary systems was first suggested by A. Toomre in 1985 (private communication, quoted by Kenyon \etal 1988 and Hartmann and Kenyon 1996), and it is supported by the numerical simulations of Bonnell and Bastien (1992) which show bursts of rapid accretion triggered by tidal interactions in a forming binary system.  However the nature of the outburst itself remains to be clarified; the idea that it is due to the rapid brightening of a self-luminous accretion disk (Kenyon \etal 1988; Hartmann and Kenyon 1996) has been questioned by Herbig \etal (2003), who argue that there is no clear spectroscopic evidence that the observed radiation from FU~Ori objects comes from disks, and suggest that these objects are better interpreted as rapidly rotating stars whose outer layers have been heated and expanded by some dynamical disturbance, as suggested by Larson (1980).  One source of such a disturbance could be an episode of rapid accretion from a disk, possibly induced by a tidal interaction with a companion star as suggested above; this is perhaps more plausible than the rotational instability  originally suggested by Larson (1980).   If a causal connection could be established among protostellar interactions, FU~Ori outbursts, and Herbig-Haro jets, this would provide strong evidence that tidal interactions play a central role in the star formation process.

\subsection{Implications for planet formation} % 6.4

   If protostellar interactions play an important role in star formation, this will clearly have important consequences for planet formation too.  Quiescent disks like the `solar nebula' in which our own planetary system is thought to have formed may in this case not be as common as has been assumed, and planet-forming disks may often be disturbed by interactions with companion stars in forming systems of stars.  Even our own solar system may have experienced significant disturbances early in its history, as is suggested by the fact that the fundamental plane of our planetary system is tilted by 8 degrees with respect to the Sun's equatorial plane; this tilt could plausibly have been caused by an encounter with another star in a young multiple system or cluster (Herbig and Terndrup 1986; Heller 1993).  If nearly all stars form in multiple systems, single stars with disks may mostly originate by ejection from such systems, as indeed happens in the detailed simulation of cluster formation by Bate \etal (2003).  Our own planetary system might then represent only a somewhat accidental particular case and might not be typical, since chaotic star formation processes may produce a large dispersion in the properties of planet-forming disks.

   Another consequence of a dynamic and chaotic picture of star formation in which circumstellar disks are often disturbed by interactions is that shocks generated by such disturbances can produce transient heating events that might account for the high-temperature inclusions observed in meteorites.  Chondrules, for example, show evidence for recurrent short heating events that reach temperatures of the order of 2000~K (Jones \etal 2000).  Shock waves in protostellar disks with speeds of several km/sec could produce temperatures of the required order for the required short times, and thus could provide a viable heating mechanism (Connolly and Love 1998; Jones \etal 2000; Desch and Connolly 2002; Susa and Nakamoto 2002).  Shocks of this strength could plausibly be produced by tidal interactions in forming systems of stars, such as the frequent interactions seen in the simulation of cluster formation by Bate \etal (2003).  The chondrules in meteorites might then bear evidence of a violent early history of our Solar system.

\section{Formation of massive stars and clusters} % 7

   The previous sections have dealt mainly with the formation of stars of low and intermediate mass, that is, stars with masses up to about 10 solar masses.  The formation of stars with larger masses clearly involves more complex processes and is less well understood, both observationally and theoretically; reviews of this subject have been given by Evans (1999), Garay and Lizano (1999), Stahler \etal (2000), and Crowther (2002).  One difference between the formation of massive stars and that of low-mass stars is that the most massive stars have masses that are much larger than the Jeans mass in the cloud cores in which they form; although these more massive cores have higher temperatures than the smaller ones in which low-mass stars form, they also have much higher densities, and the result is that the Jeans mass is not very different from that in the low-mass cores.  Thus the large cloud cores in which massive stars form might be expected to contain many smaller bound clumps, as is indeed suggested by some of the evidence (Evans 1999; Molinari \etal 2002), so these cores might form groups or clusters of stars rather than individual stars or binary systems.  Furthermore, the more massive cloud cores all contain supersonic internal turbulent motions which will generate large density fluctuations and clumpy substructure even if none was present initially, again suggesting that massive stars may form only in groups or clusters (Larson 1981).

   A second important difference is that the intense radiation emitted by a massive accreting protostar can produce feedback effects on the infalling envelope that limit or prevent continuing accretion, for example via the effects of radiative heating, radiation pressure, or ionization on the envelope (Larson and Starrfield 1971).  The most important of these feedback effects may be radiation pressure, and detailed calculations have found that the force exerted by radiation pressure in a spherical infalling envelope may exceed the force of gravity when the protostellar mass exceeds about 10~M$_\odot$ (Kahn 1974; Yorke and Kr\"ugel 1977; Wolfire and Cassinelli 1987).  Models that assume spherical infall may, as a result, not be able to account for the formation of stars with masses much larger than 10~M$_\odot$, although they may suffice for stars up to this mass (Stahler \etal 2000).  If matter is instead accreted from a circumstellar disk, a protostar may be able to build up a larger mass before radiation pressure terminates the accretion (Nakano 1989; Jijina and Adams 1996); a detailed calculation by Yorke and Sonnhalter (2002) of collapse with rotation that incorporates radiation pressure and the formation of a viscous disk shows that in this type of model the mass of an accreting protostar may reach 30~M$_\odot$ or more, although the star formation process is still very inefficient and most of the initial mass, and even most of the disk, are eventually dispersed by radiation pressure.  Therefore, it appears that even more complex models may be needed to account for the formation of the most massive stars.

\subsection{Observational evidence} % 7.1

   Because regions of massive star formation are more complex in structure and also more distant than the well-studied regions of low-mass star formation, it is more difficult to study their structure in detail and to identify individual star-forming units (Evans \etal 2002; Bally 2002); these problems are compounded by the more complex chemistry of these regions (Langer \etal 2000) and by their complex dynamics, which often includes multiple outflows.  The densities and temperatures of massive star-forming cloud cores are both considerably higher than those of low-mass cores, and they may have densities exceeding $10^7$~cm$^{-3}$ and temperatures exceeding $100$~K in some locations (Evans 1999; Garay and Lizano 1999); however, it is not clear whether the gas with these observed properties is prestellar or has already been substantially compressed and heated by prior star formation activity (Evans \etal 2002; Bally 2002).  Therefore, the most useful evidence concerning how massive stars form may be the fossil record provided by the properties of newly formed massive stars and stellar systems.  Young massive stars are almost always located in clusters or associations (Blaauw 1964, 1991; Clarke \etal 2000), which are often hierachical in structure and contain subgroups (Testi \etal 2000; Elmegreen \etal 2000); the mass of the most massive star in each subgroup or cluster tends to increase systematically with the mass of the cluster (Larson 1982, 2003; Testi \etal 1999, 2001; Garay and Lizano 1999).  The most massive stars are often centrally located in the clusters in which they form (Larson 1982; Zinnecker \etal 1993; Hillenbrand and Hartmann 1998), and this implies that these stars must have formed near the cluster center, which suggests, in turn, that accumulation processes have played an important role in their formation (Larson 1982; Bonnell and Davies 1998).  The most massive stars also have an unusually high frequency of close companions in binary or multiple systems, and these companions are themselves often massive objects (Stahler \etal 2000; Preibisch \etal 2001; Garcia and Mermilliod 2001).  The evidence thus clearly indicates that massive stars form in exceptionally dense and complex environments, typically at the centers of clusters and in close proximity to other stars which are themselves often massive.

\subsection{Possible formation processes} % 7.2

   Several authors have suggested that the formation of massive stars can be modeled with a suitable extension of the standard model for low-mass star formation that was developed by Shu \etal (1987, 1993, 1999) and was discussed above in Sections 4 and~5 (Caselli and Myers 1995; Garay and Lizano 1999; Maeder and Behrend 2002; McKee and Tan 2003).  In support of this possibility, Garay and Lizano (1999) and Garay \etal (2003) have noted that there is evidence suggesting that disks and outflows play a similar role in the formation of massive stars as they do in the formation of low-mass stars, although the existence of circumstellar disks is less well established for massive stars than for low-mass stars.  These authors also note that if the ambient gas density and hence the accretion rate are several orders of magnitude higher for massive protostars than they are for low-mass protostars, as is suggested by the evidence, then the ram pressure of the inflow may overcome radiation pressure in the infalling envelope and allow continuing accretion to occur even for luminous objects (Wolfire and Cassinelli 1987; McKee and Tan 2003).

   On the other hand, the fact that massive stars tend to form near the centers of dense clusters with many close companions suggests that protostellar interactions may play a particularly important role in the formation of massive stars and may help to drive the required high mass accretion rates (Larson 1982, 1999, 2002; Stahler \etal 2000).  An extreme type of interaction that can occur in a sufficiently dense environment is direct protostellar collisions and mergers, and such mergers have been suggested to play an important role in building up the most massive stars (Bonnell \etal 1998; Bonnell 2002).  Mergers would clearly be the most effective way of overcoming radiation pressure and other feedback effects, and they might also help to account for the high frequency of close companions of massive stars as the results of failed mergers (Zinnecker and Bate 2002).  Even though extremely high densities are required, mergers must sometimes occur because some binary systems are observed that contain stars which are almost in contact; if nature can make stars that are almost in contact, it must surely make some that come even closer together and merge.  Given the existence of the two apparently competing hypotheses, it has been debated whether massive stars form by `accretion' or by `mergers' (Crowther 2002), but the implied dichotomy between accretion and mergers is almost certainly oversimplified, and both types of processes probably play a role in the formation of massive stars.

   Important progress has come from simulations of the formation of stars with a wide range of masses in forming clusters of stars (Klessen and Burkert 2001; Bonnell and Bate 2002; Bate \etal 2003; Bonnell \etal 2003).  In these simulations, most of the stars form in subgroups at the nodes of a filamentary network, a feature that appears to be a generic result of gravitational collapse on scales much larger than the Jeans length (see also Klein \etal 2001; Balsara \etal 2001), and these subgroups eventually merge into a single large cluster.  A spectrum of protostellar masses is built up as protostars in different environments accrete mass at different rates, the more massive stars forming by faster runaway accretion processes in the denser environments and the larger subgroups (Bonnell \etal 1997).  In the simulation by Bonnell \etal (2003) of the formation of a cluster of about 400 objects in an isothermally collapsing cloud, the most massive object eventually acquires a mass of about 27~M$_\odot$, while in the simulation by Bonnell and Bate (2002) of a cluster of 1000 accreting protostars that gain mass by both accretion and mergers, the most massive object attains a mass of about 50~M$_\odot$, or 100 times the mass of a typical object.  Because the accretion rate increases with the mass of the accreting object and with the ambient density, which also increases with time as the subgroups condense and merge, the most massive objects tend to undergo a runaway growth in mass.

   The most massive object in the simulation of Bonnell and Bate (2002) forms in a compact central cluster core which undergoes a runaway increase in density because of dissipative interactions (Larson 1990a), a result somewhat analogous to the runaway growth of a central singularity in a collapsing isothermal sphere (Section~4.2).  This most massive object acquires about half of its final mass by accretion and half by mergers, although the importance of mergers may be somewhat exaggerated in this simulation by an artificially large merger cross section.  Mergers typically occur in close binary systems when they are perturbed by interaction with another protostar, causing the two protostars in the binary to merge and a new binary to be formed by the capture of the third object.  As a result of all of these complex processes, a mass spectrum resembling the Salpeter (1955) power law is built up; while there is no simple quantitative explanation for this result, a power law mass spectrum is predicted if the accretion rate increases with a power of the stellar mass (Zinnecker 1982).  More generally, a power-law mass function might be expected if the accumulation processes that build up the more massive stars are scale-free and involve no new mass scale larger than the Jeans mass (Larson 1991, 1999, 2003).  The simulations thus appear to capture, at least qualitatively, many aspects of star formation in clusters, including the formation of the most massive stars at the center, the accompanying formation of many binary and multiple systems, and the emergence of a realistic stellar mass spectrum.  The simulations also demonstrate clearly that these processes are all strongly coupled and that they involve complex dynamics and interactions between protostars and residual gas.  The formation of the most massive stars can perhaps be regarded as the culmination of all of the processes that have been discussed.

\subsection{Formation of the most massive objects} % 7.3

   A question of long-standing interest is whether there is an upper limit on stellar masses.  Although it has been suggested that feedback effects might set an upper limit of the order of 60~M$_\odot$ on the mass that an accreting protostar can attain (Larson and Starrfield 1971), these effects depend on the detailed dynamics of the star formation process, and if a protostar grows by accreting extremely dense gas or by mergers, there is no clear upper limit to the mass that it can attain.  It is also not clear from observations that there is any well-defined upper limit on stellar masses, since clusters with larger masses tend to contain more massive stars, and stars with masses up to at least 150~M$_\odot$ are found in some young clusters with masses of several times $10^4$~M$_\odot$ (Larson 2003).  The mass of the most massive star in a star-forming cloud increases systematically with the mass of the cloud (Larson 1982), and on the basis of more limited data, a similar correlation appears to hold between with the mass of the most massive star in a cluster and the mass of the cluster, this relation being approximately $M_{\rm star} \sim 1.2\,M_{\rm cluster}^{0.45}$ (Larson 2003).  Although the data on which this relation is based are sparse, the relation does not clearly terminate at any maximum mass, and this suggests that clusters with even larger masses might form even more massive stars.  In this case, it might be that some globular clusters with masses well above $10^5$~M$_\odot$ once contained even more massive stars which have long since disappeared.

   If stars with masses significantly above 150~M$_\odot$ are sometimes formed in massive clusters, this could have some important consequences.  Metal-poor stars with masses between 140 and 260~M$_\odot$ are predicted to explode completely as very energetic supernovae at the end of their lifetimes, while stars more massive than 260~M$_\odot$ are predicted to collapse completely to black holes (Heger \etal 2003).  Thus if stars with masses above 260~M$_\odot$ were ever to form at the centers of very massive clusters, the end result could be the formation of a black hole of similar mass at the cluster center.  In sufficiently dense young clusters, the runaway merging of already formed massive stars might also result in the formation of very massive stars and eventually black holes at the cluster center (Portegies Zwart and McMillan 2002).  There has been much interest in possible evidence for the existence of massive black holes at the centers of some massive globular clusters, although the evidence is so far not conclusive (Gebhardt \etal 2002; van der Marel 2003).  In addition, some ultraluminous \hbox{X-ray} sources that are possibly associated with luminous young clusters in other galaxies have been interpreted as `intermediate-mass black holes' with masses of hundreds to thousands of solar masses (Ebisuzaki \etal 2001).  If massive black holes can indeed be formed at the centers of very massive clusters, and if such massive clusters tend to be concentrated near the centers of galaxies, this could be one path toward building up supermassive black holes in galactic nuclei (Ebisuzaki \etal 2001).  All of these possibilities remain very speculative, however, and none has yet been supported by additional evidence, so it remains unclear whether exceptionally massive stars and black holes can indeed form by mechanisms like those discussed here.  Nevertheless, the importance of understanding the formation of the most massive stars is clear, and an improved understanding of the processes involved may eventually help in understanding the processes involved in the formation and growth of massive black holes.

\section{Summary} % 8

   After several decades of study, the physical conditions in nearby star-forming molecular clouds are now fairly well understood, at least for the smaller nearby dark clouds that form mostly low-mass stars.  These clouds are very cold ($\sim 10$~K) and contain small dense prestellar cores with densities above 10$^4$~H$_2$ molecules per cm$^3$ that are strongly self-gravitating and probably at an early stage of forming stars.  The origin of these dense cores is not yet well understood, but they probably form through the combined action of gravity and turbulent motions on larger scales.  Supersonic turbulence may play a role in generating the complex hierarchical structure of molecular clouds, which produces hierachical groupings of stars.  The age spans of the newly formed stars and clusters associated with molecular clouds are quite short, indicating that these clouds are transient features and that star formation is a rapid process that occurs on a dynamical timescale.

   Turbulence and magnetic fields may inhibit collapse on large scales in molecular clouds, but these effects become less important in prestellar cores where the turbulence is subsonic and magnetic fields decouple from the gas by ambipolar diffusion.  Thermal pressure is then the most important force resisting gravity, and it sets a lower limit to the sizes and masses of clumps that can collapse to form stars.  Although the original derivation by Jeans of a minimum scale for collapse was not self-consistent, its relevance is supported by analyses of the stability of various equilibrium configurations including sheets, disks, filaments, and spheres.  The observed dense cores in molecular clouds are sometimes approximated as isothermal spheres, and the stability of such spheres against collapse is governed by the Bonnor-Ebert criterion, which is dimensionally equivalent to the Jeans criterion.  For the conditions observed to exist in the densest parts of molecular clouds, the minimum unstable mass or `Jeans mass' is of the order of one solar mass, in rough agreement with the typical observed masses of the prestellar cloud cores.

   Once collapse begins, cooling by thermal emission from dust grains keeps the temperature close to 10~K while the density increases by many orders of magnitude.  The resulting nearly isothermal collapse is always highly nonuniform and is characterized by the runaway growth of a central density peak which evolves in an approximately self-similar way toward a central singularity.  As long as collapse can occur, rotation and magnetic fields do not prevent the formation a density singularity, which always develops in a qualitatively similar way.  The formation of a star can be identified mathematically with the formation of this singularity.  The importance of this result for the subsequent evolution of the system is that a very small fraction of the mass of a collapsing cloud core first attains stellar density and becomes an embryonic star, which then continues to grow in mass by the infall of matter from the surrounding envelope.

   In detail, a hydrostatic core forms when the central region becomes opaque to the thermal dust emission and its temperature rises rapidly, but this first core quickly collapses again to form a second hydrostatic core or protostar of stellar density.  This protostar initially has a very small mass of less than $10^{-2}$ solar masses and is surrounded by a centrally condensed infalling envelope that continues to fall onto it through an accretion shock at its surface.  All collapse calculations predict that the protostellar accretion rate is initially high but falls off rapidly with time as the infalling envelope is depleted.  A more idealized `standard model' that assumes an initially static protostellar envelope predicts an accretion rate that is constant in time.  In either case, a star of one solar mass is predicted to form in less than 1 million years.  During most of this time a protostar remains obscured by dust in the infalling envelope and the object is observable only at infrared wavelengths.  A young star becomes visible after accretion has ceased to be important, with predicted properties that are in reasonable agreement with the observed properties of the youngest known stars, the T~Tauri stars.

   Rotation will generally cause some of the infalling matter to form a centrifugally supported disk around an accreting protostar.  Much of the matter acquired by a star might then be accreted from such a disk, but this requires angular momentum to be removed or transported outward through the disk, and the processes involved remain poorly understood.  No completely satisfactory steady transport mechanism is known, but unsteady or episodic accretion might occur as a result of instabilities or tidal interactions.  The observed properties of the youngest stars are more variable and complex than the simplest models would predict, possibly because of variable accretion; for example, some young stars show large flareups in luminosity that could be caused by discrete accretion events, and many young objects produce sporadic jet-like outflows that are believed to be powered by episodes of rapid accretion early in the evolution of these objects.

   One likely reason for this complexity is that most stars form not in isolation but in binary or multiple systems, which in turn often form in larger groupings or clusters.  The available evidence is consistent with the possibility that {\it all} stars form in multiple systems of some kind, and that the minority of stars that now appear single have originated as escapers from such systems.  Gravitational interactions in forming systems of stars can then play an important role in the star formation process, and tidal interactions may transfer angular momentum from protostellar disks to stellar orbital motions, driving episodes of rapid accretion from disks onto forming stars.  A large fraction of jet sources have close stellar companions, and this suggests a causal connection, possibly via tidal interactions, between the presence of a companion and the launching of a jet.

   Detailed simulations of the formation of groups and clusters of stars are now possible, and they illustrate many of the features expected from the more idealized models, such as a highly nonuniform collapse, the formation of small accreting protostars with infalling envelopes or disks, and accretion at a rate that can be highly variable but generally declines with time.  Numerous binary and multiple systems are also formed in these simulations, and the dynamics of forming groups is often violent and chaotic, leading to the continuing formation and destruction of binary systems and disks.  If these simulations are representative, many protostellar disks may experience a violent early history; our solar system may itself have experienced strong disturbances early in its history which tilted the disk of our planetary system and produced strong shock heating events that generated the chondrules in meteorites.

   Some important outcomes of star formation processes that are not yet fully understood include the distribution of stellar masses, the statistical properties of binary and multiple systems, and the formation of the most massive stars.  Typical stellar masses may be determined by typical cloud properties, such as the temperature and density that control the Jeans mass, but the detailed thermal physics of the gas may also play an important role, and this has not yet been carefully studied.  The form of the mass spectrum may reflect in addition the stochastic variability that can occur in the fragmentation and accretion processes involved.  The properties of binary and multiple systems probably result from the complex and chaotic interactions that occur in forming groups or clusters of stars, which can create a large dispersion in the properties of the resulting systems.  These statistical outcomes of star formation may provide strong tests of simulations that endeavor to include all of the relevant physics; the existing simulations produce results that are qualitatively very promising, but quantitative statistical predictions remain a challenge for future work.

   The formation of the most massive stars also remains poorly understood, and it involves complicated thermal and radiation physics that have so far been studied only in idealized cases.  Additional dynamical processes such as direct stellar collisions and mergers in very dense environments may also play a role.  It is not presently clear either observationally or theoretically whether there is any upper limit to the masses with which stars can form.  If the formation of exceptionally massive stars can occur in sufficiently extreme conditions, such as might exist for example in forming galaxies, new phenomena such as the formation and growth of massive black holes might occur through an extension of the processes that have been discussed here.  Some of the processes of black hole formation and growth that have been considered in the literature are analogous to the processes of star formation and accretion that have been discussed in this review.  Thus further study of these processes may have broader ramifications in astrophysics.  As was originally foreseen by Newton, gravity can collect much of the diffuse matter in the universe into compact objects, but the ensuing interactions between gravity and other forces can produce a wide array of astrophysical phenomena that can include some of the most extreme and energetic phenomena observed in the universe.

\References

\item[] Abt H A 1983 {\it Ann.\ Rev.\ Astron.\ Astrophys.} {\bf 21}
  343--372

\item[] Adams F C and Fatuzzo M 1996 {\it Astrophys.\ J.} {\bf 464}
  256--271

\item[] Adams F C and Lin D N C 1993 {\it Protostars and Planets III}
  ed E~H Levy and J~I Lunine (Tucson:\ Univ.\ of Arizona Press)
  pp~721--748

\item[] Aikawa Y, Ohashi N and Herbst E 2003 {\it Astrophys.\ J.}
  in press (astro-ph/0305054)

\item[] Andr\'e P, Motte F, Bacmann A and Belloche A 1999 {\it Star
  Formation 1999} ed T Nakamoto (Nobeyama:\ Nobeyama Radio Observatory)
  pp~145--152

\item[] Andr\'e P, Ward-Thompson D  and Barsony M 2000 {\it Protostars
  and Planets IV} ed V Mannings \etal (Tucson:\ Univ.\ of Arizona
  Press) pp~59--96

\item[] Andr\'e P, Bouwman J, Belloche A and Hennebelle P 2003 {\it
  Chemistry as a Diagnostic of Star Formation} ed C~L Curry and M Fich 
  (Ottawa: NRC Press) in press (astro-ph/0212492)

\item[] Appenzeller I and Tscharnuter W 1975 {\it Astron.\ Astrophys.}
  {\bf 40} 397--399

\item[] Arce H G and Goodman A~A, 2002 {\it Astrophys.\ J.} {\bf 575}
  911--927

\item[] Arons J and Max C E 1975 {\it Astrophys.\ J.} {\bf 196} L77--81

\item[] Balbus S A and Hawley J F 1998 {\it Rev.\ Mod.\ Phys.} {\bf 70}
  1--53

\item[] Ballesteros-Paredes J 2003 {\it From Observations to
  Self-Consistent Modeling of the Interstellar Medium,} {\it Astrophys.\
  Space Sci.} in press (astro-ph/0212580)

\item[] Ballesteros-Paredes J, Hartmann L and V\'azquez-Semadeni E 1999
  {\it Astrophys.\ J.} {\bf 527} 285--297

\item[] Bally J 2002 {\it Hot Star Workshop III: The Earliest Stages of
  Massive Star Birth (ASP Conf.\ Ser.\ 267)} ed P~A Crowther
  (San Francisco:\ Astron.\ Soc.\ Pacific) pp~219--233

\item[] Balsara D, Ward-Thompson D and Crutcher R M 2001 {\it Mon.\
  Not.\ R. Astron.\ Soc.} {\bf 327} 715--720

\item[] Basu S 1997 {\it Astrophys.\ J.} {\bf 485} 240--253

\item[] Basu S 1998 {\it Astrophys.\ J.} {\bf 509} 229--237

\item[] Basu S and Mouschovias T Ch 1994 {\it Astrophys.\ J.} {\bf 432}
  720--741

\item[] Basu S and Mouschovias T Ch 1995a {\it Astrophys.\ J.} {\bf 452}
  386--400

\item[] Basu S and Mouschovias T Ch 1995b {\it Astrophys.\ J.} {\bf 453}
  271--283

\item[] Basu S and Murali C 2001 {\it Astrophys.\ J.} {\bf 551} 743--746

\item[] Bate M R 1998 {\it Astrophys.\ J.} {\bf 508} L95--98

\item[] Bate M R 2000 {\it Mon.\ Not.\ R. Astron.\ Soc.} {\bf 314}
  33--53

\item[] Bate M R and Bonnell I A 1997 {\it Mon.\ Not.\ R. Astron.\ Soc.}
  {\bf 285} 33--48

\item[] Bate M R, Bonnell I A and Bromm V 2002a {\it Mon.\ Not.\ R.
  Astron.\ Soc.} {\bf 332} L65--68

\item[] Bate M R, Bonnell I A and Bromm V 2002b {\it Mon.\ Not.\ R.
  Astron.\ Soc.} {\bf 336} 705--713

\item[] Bate M R, Ogilvie G I, Lubow S H and Pringle J E 2002c {\it
  Mon.\ Not.\ R.\ Astron.\ Soc.} {\bf 332} 575--600

\item[] Bate M R, Bonnell I A and Bromm V 2003 {\it Mon.\ Not.\ R.
  Astron.\ Soc.} {\bf 339} 577--599

\item[] Beckwith S V W 1999 {\it The Origin of Stars and Planetary
  Systems} ed C~J Lada and N~D Kylafis (Dordrecht: Kluwer) pp~579--611

\item[] Belloche A, Andr\'e P, Despois D and Blinder S 2002 {\it
  Astron.\ Astrophys.} {\bf 393} 927--947

\item[] Binney J and Tremaine S 1987 {\it Galactic Dynamics}
  (Princeton: Princeton Univ.\ Press)  

\item[] Blaauw A 1964 {\it Ann.\ Rev.\ Astron.\ Astrophys.} {\bf 2}
  213--246

\item[] Blaauw A 1991 {\it The Physics of Star Formation and Early
  Stellar Evolution} ed C~J Lada and N~D Kylafis (Dordrecht:\ Kluwer)
  pp~125--154

\item[] Black D C and Bodenheimer P 1976 {\it Astrophys.\ J.} {\bf 206}
  138--149

\item[] Black D C and Matthews M S 1985 (eds) {\it Protostars and
  Planets II} (Tucson:\  Univ.\ of Arizona Press)

\item[] Blitz L 1993 {\it Protostars and Planets III} ed E~H Levy and
  J~I Lunine (Tucson:\  Univ.\ of Arizona Press) pp~125--161

\item[] Blitz L and Williams J P 1999 {\it The Origin of Stars and
  Planetary Systems} ed C~J Lada and N~D Kylafis (Dordrecht:\ Kluwer)
  pp~3--28

\item[] Blondin J M 2000 {\it New Astron.} {\bf 5} 53--68

\item[] Bodenheimer P 1968 {\it Astrophys.\ J.} {\bf 153} 483--494

\item[] Bodenheimer P 1972 {\it Rep.\ Prog.\ Phys.} {\bf 35} 1--54

\item[] Bodenheimer P 1978 {\it Astrophys.\ J.} {\bf 224} 488--496

\item[] Bodenheimer P 1992 {\it Star Formation in Stellar Systems} ed
  G Tenorio-Tagle \etal (Cambridge: Cambridge Univ.\ Press) pp~1--65

\item[] Bodenheimer P 1995 {\it Ann.\ Rev.\ Astron.\ Astrophys.}
  {\bf 33} 199--238

\item[] Bodenheimer P and Sweigart A 1968 {\it Astrophys.\ J.} {\bf 152}
  515--522

\item[] Bodenheimer P, Ruzmaikina T and Mathieu R D 1993 {\it Protostars
  and Planets III} ed E~H Levy and J~I Lunine (Tucson:\ Univ.\ of
  Arizona Press) pp~367--404

\item[] Bodenheimer P, Burkert A, Klein R I and Boss A P 2000
  {\it Protostars and Planets IV} ed V Mannings \etal (Tucson:\ Univ.\
  of Arizona Press) pp~675--701

\item[] Boffin H M J 2001 {\it Astrotomography: Indirect Imaging
  Methods in Observational Astronomy (Lecture Notes in Physics 573)}
  ed H~M~J Boffin \etal (Berlin: Springer) pp~69--87

\item[] Bok B J 1948 {\it Harvard Observatory Monographs, No.\ 7:
  Centennial Symposia} ed M W Mayall (Cambridge, Mass.: Harvard College
  Observatory) pp~53--72

\item[] Bok B J, Cordwell C S and Cromwell R H 1971 {\it Dark Nebulae,
  Globules, and Protostars} ed B T Lynds (Tucson:\ Univ.\ of Arizona
  Press) pp~33-55

\item[] Bonnell I A 1999 {\it The Origin of Stars and Planetary Systems}
   ed C~J Lada and N~D Kylafis (Dordrecht: Kluwer) pp~479--513

\item[] Bonnell I A 2002 {\it Hot Star Workshop III: The Earliest Stages
  of Massive Star Birth (ASP Conf.\ Ser.\ 267)} ed P~A Crowther
  (San Francisco: Astron.\ Soc.\ Pacific) pp~193--208

\item[] Bonnell I A and Bastien P 1992 {\it Astrophys.\ J.} {\bf 401}
  L31--34

\item[] Bonnell I A and Bate M R 2002, {\it Mon.\ Not.\ R. Astron.\
  Soc.} {\bf 336} 659--669

\item[] Bonnell I A and Davies M B 1998 {\it Mon.\ Not.\ R. Astron.\
  Soc.} {\bf 295} 691--698

\item[] Bonnell I A, Bate M R, Clarke C J and Pringle J E 1997 {\it
  Mon.\ Not.\ R. Astron.\ Soc.} {\bf 285} 201--208

\item[] Bonnell I A, Bate M R and Zinnecker H 1998 {\it Mon.\ Not.\ R.
  Astron.\ Soc.} {\bf 298} 93--102

\item[] Bonnell I A, Bate M R and Vine S G 2003 {\it Mon.\ Not.\ R.
  Astron.\ Soc.} {\bf 343} 413--418

\item[] Bonnor W B 1956 {\it Mon.\ Not.\ R. Astron.\ Soc.} {\it 116}
  351--359

\item[] Boss A P 1990 {\it Physical Processes in Fragmentation and Star
  Formation} ed R Capuzzo-Dolcetta \etal (Dordrecht:\ Kluwer)
  pp~279--292

\item[] Boss A P 1993a {\it The Realm of Interacting Binary Stars} ed
  J Sahade \etal (Dordrecht:\ Kluwer) pp~355--379

\item[] Boss A P 1993b {\it Astrophys.\ J.} {\bf 410} 157--167

\item[] Boss A P 2001 {\it Astrophys.\ J.} {\bf 563} 367--373

\item[] Boss A P 2002 {\it Astrophys.\ J.} {\bf 576} 462--472

\item[] Bourke T L, Myers P C, Robinson G and Hyland A R 2001 {\it
  Astrophys.\ J.} {\bf 554} 916--932

\item[] Burkert A, Bate M R and Bodenheimer P 1997 {\it Mon.\ Not.\ R.
  Astron.\ Soc.} {\bf 289} 497--504

\item[] Burkert A and Bodenheimer P 2000 {\it Astrophys.\ J.} {\bf 543}
  822--830

\item[] Calvet N, Hartmann L and Strom S E 2000 {\it Protostars and
  Planets IV} ed V Mannings \etal (Tucson:\ Univ.\ of Arizona Press)
  pp~377--399

\item[] Caselli P and Myers P C 1995 {\it Astrophys.\ J.} {\bf 446}
  665--686

\item[] Cernicharo J 1991 {\it The Physics of Star Formation and Early
  Stellar Evolution} ed C~J Lada and N~D Kylafis (Dordrecht:\ Kluwer)
  pp~287--328

\item[] Chandrasekhar S and Fermi E 1953 {\it Astrophys.\ J.} {\bf 118}
  116--141

\item[] Ciolek G E and Basu S 2000 {\it Astrophys.\ J.} {\bf 529}
  925--931

\item[] Ciolek G E and Basu S 2001 {\it Astrophys.\ J.} {\bf 547}
  272--279

\item[] Ciolek G E and K\"onigl A 1998 {\it Astrophys.\ J.} {\bf 504}
  257--279

\item[] Clarke C J, Bonnell I A and Hillenbrand L A 2000 {\it Protostars
  and Planets IV} ed V Mannings \etal (Tucson:\ Univ.\ of Arizona Press)
  pp~151--177

\item[] Cohen M and Kuhi L V 1979 {\it Astrophys.\ J. Suppl.} {\bf 41}
  743--843

\item[] Connolly H C and Love S G 1998 {\it Science} {\bf 280} 62--67

\item[] Contopoulos I, Ciolek G E and K\"onigl A 1998 {\it Astrophys.\
  J.} {\bf 504} 247--256

\item[] Crowther P A 2002 (ed) {\it Hot Star Workshop III: The Earliest
  Stages of Massive Star Birth (ASP Conf.\ Ser.\ 267)} (San
  Francisco: Astron.\ Soc.\ Pacific)

\item[] Crutcher R M 1999 {\it Astrophys.\ J.} {\bf 520} 706--713

\item[] Curry C L 2000 {\it Astrophys.\ J.} {\bf 541} 831--840

\item[] Curry C L 2002 {\it Astrophys.\ J.} {\bf 576} 849--859

\item[] Desch S J and Connolly H C 2002 {\it Meteoritics \& Planetary
  Science} {\bf 37} 183--207

\item[] Dickman R L 1985 {\it Protostars and Planets II} ed D~C Black
  and M~S Matthews (Tucson:\ Univ.\ of Arizona Press) pp~150--174

\item[] Dopita M A 1978 {\it Astrophys.\ J. Suppl.} {\bf 37} 117--144

\item[] Duquennoy A and Mayor M 1991 {\it Astron.\ Astrophys.} {\bf 248}
  485--524

\item[] Ebert R 1957 {\it Zeitschrift f\"ur Astrophysik} {\bf 42}
  263--272

\item[] Ebisuzaki T, Makino J, Tsuru T, Funato Y, Portegies Zwart S,
  Hut P, McMillan S, Matsushita S, Matsumoto H and Kawabe R 2001
  {\it Astrophys.\ J.} {\bf 562} L19--L22

\item[] Edwards S, Ray T and Mundt R 1993 {\it Protostars and Planets
  III} ed E~H Levy and J~I Lunine (Tucson:\ Univ.\ of Arizona Press)
  pp~567--602

\item[] Eisl\"offel J, Mundt R, Ray T P and Rodriguez L P 2000 {\it
  Protostars and Planets IV} ed V Mannings \etal (Tucson:\ Univ.\ of
  Arizona Press) pp~815--840

\item[] Elmegreen B G 1985 {\it Protostars and Planets II} ed D~C Black
  and M~S Matthews (Tucson:\ Univ.\ of Arizona Press) pp~33--58

\item[] Elmegreen B G 1992 {\it Star Formation in Stellar Systems} ed
  G Tenorio-Tagle \etal (Cambridge: Cambridge Univ.\ Press) pp~381--478

\item[] Elmegreen B G 1993 {\it Protostars and Planets III} ed E~H Levy
  and J~I Lunine (Tucson:\ Univ.\ of Arizona Press) pp~97--124

\item[] Elmegreen B G 1997 {\it Astrophys.\ J.} {\bf 486} 944--954

\item[] Elmegreen B G 1999 {\it Astrophys.\ J.} {\bf 515} 323--336

\item[] Elmegreen B G 2000 {\it Astrophys.\ J.} {\bf 530} 277--281

\item[] Elmegreen B G and Elmegreen D M 1978 {\it Astrophys.\ J.} {\bf
  220} 1051--1062

\item[] Elmegreen B G, Efremov Y, Pudritz R E and Zinnecker H 2000
  {\it Protostars and Planets IV} ed V Mannings \etal (Tucson:\ Univ.\
  of Arizona Press) pp~179--215

\item[] Evans N J 1999 {\it Ann.\ Rev.\ Astron.\ Astrophys.} {\bf 37}
  311--362

\item[] Evans N J, Shirley Y L, Mueller K E and Knez C 2002 {\it Hot
  Star Workshop III: The Earliest Stages of Massive Star Birth (ASP
  Conf.\ Ser.\ 267)} ed P~A Crowther (San Francisco: Astron.\
  Soc.\ Pacific) pp~17--31

\item[] Falgarone E and Phillips T G 1991 {\it Fragmentation of
  Molecular Clouds and Star Formation (IAU Symp.\ 147)} ed E Falgarone
  \etal (Dordrecht:\ Kluwer) pp~119--136

\item[] Falgarone E, Phillips T G and Walker C K 1991 {\it Astrophys.\
  J.} {\bf 378} 186--201

\item[] Falgarone E, Puget J-L and P\'erault M 1992 {\it Astron.\
  Astrophys.} {\bf 257} 715--730

\item[] Foster P N and Chevalier R A 1993 {\it Astrophys.\ J.} {\bf 416}
  303--311

\item[] Franco J, Shore S N and Tenorio-Tagle G 1994 {\it Astrophys.\
  J.} {\bf 436} 795--799

\item[] Fromang S, Terquem C and Balbus S A 2002 {\it Mon.\ Not.\ R.
  Astron.\ Soc.} {\bf 329} 18--28

\item[] Fukui Y, Iwata T, Mizuno A, Bally J and Lane A P 1993 {\it
  Protostars and Planets III} ed E~H Levy and J~I Lunine (Tucson:\
  Univ.\ of Arizona Press) pp~603--639

\item[] Galli D, Shu F H, Laughlin G and Lizano S 2001 {\it Astrophys.\
  J.} {\bf 551} 367--386

\item[] Gammie C F 1996 {\it Astrophys.\ J.} {\bf 457} 355--362 

\item[] Gammie C F 2001 {\it Astrophys.\ J.} {\bf 553} 174--183

\item[] Gammie C F, Lin Y-T, Stone J M and Ostriker E C 2003 {\it
  Astrophys.\ J.} {\bf 592} 203--216

\item[] Garay G and Lizano S 1999 {\it Publ.\ Astron.\ Soc.\ Pacific}
  {\bf 111} 1049--1087

\item[] Garay G, Brooks K J, Mardones D and Norris R P 2003 {\it
  Astrophys.\ J.} {\bf 587} 739--747 

\item[] Garcia B and Mermilliod J C 2001 {\it Astron.\ Astrophys.}
  {\bf 368} 122--136

\item[] Gebhardt K, Rich R M and Ho L C 2002 {\it Astrophys.\ J.}
  {\bf 578} L41--L45

\item[] Gehrels T 1978 (ed) {\it Protostars and Planets} (Tucson:\
  Univ.\ of Arizona Press)

\item[] Gilden D L 1984 {\it Astrophys.\ J.} {\bf 283} 679--686

\item[] Goldman I 2000 {\it Astrophys.\ J.} {\bf 541} 701--706

\item[] Goldreich P and Lynden-Bell D 1965a {\it Mon.\ Not.\ R.
  Astron.\ Soc.} {\bf 130} 97--124

\item[] Goldreich P and Lynden-Bell D 1965b {\it Mon.\ Not.\ R.
  Astron.\ Soc.} {\bf 130} 125--158

\item[] Goldreich P and Tremaine S 1982 {\it Ann.\ Rev.\ Astron.\
  Astrophys.} {\bf 20} 249--283

\item[] Gomez M, Hartmann L, Kenyon S J and Hewett R 1993 {\it Astron.\
  J.} {\bf 105} 1927--1937

\item[] Goodman J and Rafikov R~R 2001 {\it Astrophys.\ J.} {\bf 552}
  793--802

\item[] Goodman A~A, Bastien P, Menard F and Myers P C 1990 {\it
  Astrophys.\ J.} {\bf 359} 363--377

\item[] Goodman A~A, Benson P J, Fuller G A and Myers P C 1993 {\it
  Astrophys.\ J.} {\bf 406} 528--547

\item[] Goodman A~A, Barranco J A, Wilner D J and Heyer M H 1998
  {\it Astrophys.\ J.} {\bf 504} 223--246

\item[] Hanawa T and Matsumoto T 1999 {\it Astrophys.\ J.} {\bf 521}
  703--707

\item[] Hanawa T and Matsumoto T 2000 {\it Publ.\ Astron.\ Soc.\ Japan}
  {\bf 52} 241--247

\item[] Hartmann L 1998 {\it Accretion Processes in Star Formation}
  (Cambridge: Cambridge Univ.\ Press)

\item[] Hartmann L 2001 {\it Astron.\ J.} {\bf 121} 1030--1039

\item[] Hartmann L 2002 {\it Astrophys.\ J.} {\bf 578} 914--924

\item[] Hartmann L 2003 {\it Astrophys.\ J.} {\bf 585} 398--405

\item[] Hartmann L and Kenyon S 1996 {\it Ann.\ Rev.\ Astron.\
  Astrophys.} {\bf 34} 207--240

\item[] Hartmann L, Kenyon S and Hartigan P 1993 {\it Protostars and
  Planets III} ed E~H Levy and J~I Lunine (Tucson: Univ.\ of Arizona
  Press) pp~497--518

\item[] Hartmann L, Cassen P and Kenyon S J 1997 {\it Astrophys.\ J.}
  {\bf 475} 770--785 

\item[] Hartmann L, Calvet N, Gullbring E and D'Alessio P 1998 {\it
  Astrophys.\ J.} {\bf 495} 385--400

\item[] Hartmann L, Ballesteros-Paredes J and Bergin E A 2001 {\it
  Astrophys.\ J.} {\bf 562} 852--868

\item[] Hayashi C 1966 {\it Ann. Rev.\ Astron.\ Astrophys.} {\bf 4}
  171--192

\item[] Hayashi C and Nakano T 1965 {\it Prog.\ Theor.\ Phys.} {\bf 34}
  754--775

\item[] Hayashi C, Hoshi R and Sugimoto D 1962 {\it Prog.\ Theor.\
  Phys.\ Suppl.} No.\ {\bf 22} 1--183

\item[] Hayashi C, Narita S and Miyama S M 1982 {\it Prog.\ Theor.\
  Phys.} {\bf 68} 1949

\item[] Heacox W D 1998 {\it Astron.\ J.} {\bf 115} 325--337

\item[] Heacox W D 2000 {\it Birth and Evolution of Binary Stars
  (Poster Proceedings of IAU Symp.\ 200)} ed B Reipurth and H Zinnecker
  (Potsdam: Astrophys.\ Inst.\ Potsdam) pp~208--210

\item[] Heger A, Fryer C L, Woosley S E, Langer N and Hartmann D H 2003
  {\it Astrophys.\ J.} {\bf 591} 288--300

\item[] Heiles C, Goodman A~A, McKee C F and Zweibel E G 1993 {\it
  Protostars and Planets III} ed E~H Levy and J~I  Lunine (Tucson:\
  Univ.\ of Arizona Press) pp~279--326

\item[] Heintz W D 1969 {\it J. R. Astron.\ Soc.\ Canada} {\bf 63}
  275--298

\item[] Heitsch F, Mac Low M-M and Klessen R S 2001 {\it Astrophys.\
  J.} {\bf 547} 280--91

\item[] Heller C H 1993 {\it Astrophys.\ J.} {\bf 408} 337--346

\item[] Heller C H 1995 {\it Astrophys.\ J.} {\bf 455} 252--259

\item[] Herbig G H 1962 {\it Advances in Astronomy and Astrophysics}
  {\bf 1} 47--103

\item[] Herbig G H 1977 {\it Astrophys.\ J.} {\bf 217} 693--715

\item[] Herbig G H and Terndrup D M 1986 {\it Astrophys.\ J.} {\bf 307}
  609--618

\item[] Herbig G H, Petrov P~P and Duemmler R 2003 {\it Astrophys.\ J.}
  {\bf 595} in press (astro-ph/0306559) 

\item[] Herbst E 1994 {\it The Structure and Content of Molecular
  Clouds (Lecture Notes in Physics 439)} ed T~L Wilson and K~J
  Johnston (Berlin: Springer) pp~29--54

\item[] Hillenbrand L A and Hartmann L W 1998 {\it Astrophys.\ J.}
  {\bf 492} 540--553

\item[] Hoyle F 1953 {\it Astrophys.\ J.} {\bf 118} 513--528

\item[] Hoyle F and Lyttleton R A 1939 {\it Proc.\ Camb.\ Phil.\ Soc.}
  {\bf 35} 592--609

\item[] Hunter C 1964 {\it Astrophys.\ J.} {\bf 139} 570--586

\item[] Hunter C 1977 {\it Astrophys.\ J.} {\bf 218} 834--845

\item[] Indebetouw R and Zweibel E G 2000 {\it Astrophys.\ J.} {\bf 532}
  361--376

\item[] Jeans J H 1902 {Phil.\ Trans.\ R. Soc.} {\bf 199} 49

\item[] Jeans J H 1929 {\it Astronomy and Cosmogony} (Cambridge:
  Cambridge Univ.\ Press; reprinted by Dover, New York, 1961)

\item[] Jijina J and Adams F C 1996 {\it Astrophys.\ J.} {\bf 462}
  874--887

\item[] Johnstone D, Wilson C D, Moriarty-Schieven G, Joncas G, Smith G,
  Gregersen E and Fich M 2000 {\it Astrophys.\ J.} {\bf 545} 327--339

\item[] Johnstone D, Fich M, Mitchell G F and Moriarty-Schieven G 2001
  {\it Astrophys.\ J.} {\bf 559} 307--317

\item[] Jones R H, Lee T, Connolly H C, Love S G and Shang H 2000 {\it
  Protostars and Planets IV} ed V Mannings \etal (Tucson: Univ.\ of
  Arizona Press) pp~927--962 

\item[] Kahn F D 1974 {\it Astron.\ Astrophys.} {\bf 37} 149--162

\item[] Kennicutt R C 1998 {\it Ann.\ Rev.\ Astron.\ Astrophys.}
  {\bf 36} 189--231

\item[] Kenyon S J and Hartmann L 1995 {\it Astrophys.\ J. Suppl.}
  {\bf 101} 117--171

\item[] Kenyon S J, Hartmann L and Hewett R 1988 {\it Astrophys.\ J.}
  {\bf 325} 231--251

\item[] Klein R I, Fisher R and McKee C F 2001 {\it The Formation of
  Binary Stars (IAU Symp.\ 200)} ed H Zinnecker and R D Mathieu
  (San Francisco: Astron.\ Soc.\ Pacific) pp~361--370

\item[] Klessen R S 2001a {\it Astrophys.\ J.} {\bf 550} L77--80

\item[] Klessen R S 2001b {\it Astrophys.\ J.} {\bf 556} 837--846

\item[] Klessen R S and Burkert A 2001 {\it Astrophys.\ J.} {\bf 549}
  386--401

\item[] Klessen R S, Heitsch F and Mac Low M-M 2000 {\it Astrophys.\
  J.} {\bf 535} 887--906

\item[] K\"onigl A and Pudritz R E 2000 {\it Protostars and Planets IV}
  ed V Mannings \etal (Tucson:\ Univ.\ of Arizona Press) pp~759--787

\item[] K\"onigl A and Ruden S P 1993 {\it Protostars and Planets III}
  ed E~H Levy and J~I Lunine (Tucson:\ Univ.\ of Arizona Press)
  pp~641--687

\item[] Kroupa P 2001 {\it Mon.\ Not.\ R. Astron.\ Soc.} {\bf 322}
  231--246

\item[] Kroupa P 2002 {\it Science} {\bf 295} 82--91

\item[] Kroupa P and Burkert A 2001 {\it Astrophys.\ J.} {\bf 555}
  945--949

\item[] Lada C J 1991 {\it The Physics of Star Formation and Early
  Stellar Evolution} ed C~J Lada and N~D Kylafis (Dordrecht: Kluwer)
  pp~329--363

\item[] Lada C J and Kylafis N D 1991 (eds) {\it The Physics of Star
  Formation and Early Stellar Evolution} (Dordrecht:\ Kluwer)

\item[] Lada C J and Kylafis N D 1999 (eds) {\it The Origin of Stars
  and Planetary Systems} (Dordrecht:\ Kluwer)

\item[] Lada E A, Strom K M and Myers P C 1993 {\it Protostars and
  Planets III} ed E~H Lunine and J~I Levy (Tucson:\ Univ.\ of Arizona
  Press) pp~245--277

\item[] Lai D 2000 {\it Astrophys.\ J.} {\bf 540} 946--961

\item[] Langer W D, van Dishoeck E F, Bergin E A, Blake G A, Tielens
  A~G~G~M, Velusamy T and Whittet D~C~B 2000 {\it Protostars and
  Planets IV} ed V Mannings \etal (Tucson:\ Univ.\ of Arizona Press)
  pp~29--57

\item[] Larson R B 1969 {\it Mon.\ Not.\ R. Astron.\ Soc.} {\bf 145}
  271--295

\item[] Larson R B 1972a {\it Mon.\ Not.\ R. Astron.\ Soc.} {\bf 156}
  437--458

\item[] Larson R B 1972b {\it Mon.\ Not.\ R. Astron.\ Soc.} {\bf 157}
  121--145

\item[] Larson R B 1973 {\it Fundam.\ Cosmic Phys.} {\bf 1} 1-70

\item[] Larson R B 1978 {\it Mon.\ Not.\ R. Astron.\ Soc.} {\bf 84}
  69--85

\item[] Larson R B 1979 {\it Mon.\ Not.\ R. Astron.\ Soc.} {\bf 186}
  479-490

\item[] Larson R B 1980 {\it Mon.\ Not.\ R. Astron.\ Soc.} {\bf 190}
  321--335

\item[] Larson R B 1981 {\it Mon.\ Not.\ R. Astron.\ Soc.} {\bf 194}
  809--826

\item[] Larson R B 1982 {\it Mon.\ Not.\ R. Astron.\ Soc.} {\bf 200}
  159--174

\item[] Larson R B 1984 {\it Mon.\ Not.\ R. Astron.\ Soc.} {\bf 206}
  197--207

\item[] Larson R B 1985 {\it Mon.\ Not.\ R. Astron.\ Soc.} {\bf 214}
  379--398

\item[] Larson R B 1988 {\it Galactic and Extragalactic Star Formation}
  ed R~E Pudritz and M Fich (Dordrecht:\ Kluwer) pp~459--474

\item[] Larson R B 1989 {\it The Formation and Evolution of Planetary
  Systems} ed H~A Weaver and L Danly (Cambridge: Cambridge Univ.\
  Press) pp~31--54

\item[] Larson R B 1990a {\it Physical Processes in Fragmentation and
  Star Formation} ed R Capuzzo-Dolcetta \etal (Dordrecht:\ Kluwer)
  pp~389--400

\item[] Larson R B 1990b {\it Mon.\ Not.\ R. Astron.\ Soc.} {\bf 243}
  588--592

\item[] Larson R B 1991 {\it Fragmentation of Molecular Clouds and Star
  Formation (IAU Symp.\ 147)} ed E Falgarone, F Boulanger and
  G Duvert (Dordrecht: Kluwer) pp~261--273

\item[] Larson R B 1992 {\it Star Formation in Stellar Systems} ed
  G Tenorio-Tagle \etal (Cambridge: Cambridge Univ.\ Press) pp~125--190

\item[] Larson R B 1994 {\it The Structure and Content of Molecular
  Clouds (Lecture Notes in Physics 439)} ed T~L Wilson and K~J
  Johnston (Berlin: Springer) pp~13--28

\item[] Larson R B 1995 {\it Mon.\ Not.\ R. Astron.\ Soc.} {\bf 272}
  213--220

\item[] Larson R B 1996 {\it The Interplay Between Massive Star
  Formation, the ISM and Galaxy Evolution} ed D Kunth \etal (Gif sur
  Yvette: Editions Fronti\`eres) pp~3--16

\item[] Larson R B 1997 {\it Structure and Evolution of Stellar
  Systems} ed T~A Agekian \etal (St.\ Petersburg, Russia: St.\
  Petersburg State University) pp~48--62
  (http://www.astro.yale.edu/larson/papers/\\Petrozavodsk95.ps)

\item[] Larson R B 1999 {\it Star Formation 1999} ed T Nakamoto
  (Nobeyama: Nobeyama Radio Observatory) pp~336--340

\item[] Larson R B 2001  {\it The Formation of Binary Stars (IAU Symp.\
  200)} ed H Zinnecker and R~D Mathieu (San Francisco: Astron.\ Soc.\
  Pacific) pp~93--106

\item[] Larson R B 2002 {\it Mon.\ Not.\ R. Astron.\ Soc.} {\bf 332}
  155--164

\item[] Larson R B 2003 {\it Galactic Star Formation Across the Stellar
  Mass Spectrum (ASP Conf.\ Ser.\ 287)} ed J~M De Buizer and N~S
  van der Bliek (San Francisco: Astron.\ Soc.\ Pacific) pp~65--80

\item[] Larson R B and Starrfield S 1971 {\it Astron.\ Astrophys.}
  {\bf 13} 190--197

\item[] Laughlin G and R\'o\.zyczka M 1996 {\it Astrophys.\ J.}
  {\bf 456} 279--291

\item[] Leisawitz D, Bash F N and Thaddeus P 1989 {\it Astrophys.\ J.
  Suppl.} {\bf 70} 731--812

\item[] Leung C M 1985 {\it Protostars and Planets II} ed D~C Black
  and M~S Matthews (Tucson:\ Univ. of Arizona Press) pp~104--136

\item[] Levy E H and Lunine J I 1993 (eds) {\it Protostars and Planets
  III} (Tucson:\ Univ.\ of Arizona Press)

\item[] Li Z-Y and Shu F H 1996 {\it Astrophys.\ J.} {\bf 472} 211--224

\item[] Lin C~C, Mestel L and Shu F H 1965 {\it Astrophys.\ J.}
  {\bf 142} 1431--1446

\item[] Lin D N C and Papaloizou J C B 1993 {\it Protostars and Planets
  III} ed E~H Levy and J~I Lunine (Tucson:\ Univ.\ of Arizona Press)
  pp~749--835

\item[] Lombardi M and Bertin G 2001 {Astron.\ Astrophys.} {\bf 375}
  1091--1099

\item[] Looney L W, Mundy L G and Welch W J 2000 {\it Astrophys.\ J.}
  {\bf 529} 477--498

\item[] Low C and Lynden-Bell D 1976 {\it Mon.\ Not.\ R. Astron.\ Soc.}
  {\bf 176} 367--390

\item[] Lubow S H and Artymowicz P 2000 {\it Protostars and Planets IV}
  ed V Mannings \etal (Tucson:\ Univ.\ of Arizona Press) pp~731--755

\item[] Luhman K L and Rieke G H 1999 {\it Astrophys.\ J.} {\bf 525}
  440--465

\item[] Lynden-Bell D and Pringle J E 1974 {\it Mon.\ Not.\ R. Astron.\
  Soc.} {\bf 168} 603--637

\item[] Mac Low M-M 2003 {\it Turbulence and Magnetic Fields in
  Astrophysics (Lecture Notes in Physics 614)} ed E Falgarone and
  T Passot (Berlin: Springer) pp~182--212

\item[] Mac Low M-M and Klessen R S 2003 {\it Rev.\ Mod.\ Phys.}
  in press (astro-ph/0301093)

\item[] Mac Low M-M, Klessen R S, Burkert A and Smith M D 1998 {\it
  Phys.\ Rev.\ Letters} {\bf 80} 2754--2757

\item[] Maeder A and Behrend R 2002 {\it Hot Star Workshop III: The
  Earliest Stages of Massive Star Birth (ASP Conf.\ Ser.\ 267)}
  ed P~A Crowther (San Francisco:\ Astron.\ Soc.\ Pacific) pp~179--192

\item[] Makita M, Miyawaki K and Matsuda T 2000 {\it Mon.\ Not.\ R.
  Astron.\ Soc.} {\bf 316} 906--916

\item[] Mannings V, Boss A P and Russell S~S 2000 (eds) {\it Protostars
  and Planets IV} (Tucson:\ Univ.\ of Arizona Press)

\item[] Masunaga H and Inutsuka S 2000 {\it Astrophys.\ J.} {\bf 531}
  350--365

\item[] Mathieu R D 1994 {\it Ann.\ Rev.\ Astron.\ Astrophys.} {\bf 32}
  465--530

\item[] Matsuda T, Makita M, Fujiwara H, Nagae T, Haraguchi K, Hayashi
  E and Boffin H~M~J 2000 {\it Astrophys.\ and Space Sci.} {\bf 274}
  259--273

\item[] Matsumoto T and Hanawa T 2003 {\it Astrophys.\ J.} in press
  (astro-ph/0306121)

\item[] Matsumoto T, Hanawa T and Nakamura F 1997 {\it Astrophys.\ J.}
  {\bf 478} 569-584

\item[] Matzner C D 2002 {\it Astrophys.\ J.} {\bf 566} 302--314

\item[] McKee C F 1999 {\it The Origin of Stars and Planetary Systems}
  ed C~J Lada and N~D Kylafis (Dordrecht:\ Kluwer) pp~29--66

\item[] McKee C F and Tan J C 2003 {\it Astrophys.\ J.} {\bf 585}
  850--871

\item[] McKee C F and Zweibel E G 1995 {\it Astrophys.\ J.} {\bf 440}
  686--696

\item[] McKee C F, Storey J W V, Watson D M and Green S 1982 {\it
  Astrophys.\ J.} {\bf 259} 647--656

\item[] McKee C F, Zweibel E G, Goodman A~A and Heiles C 1993 {\it
  Protostars and Planets III} ed E~H Levy and J~I Lunine (Tucson:
  Univ.\ of Arizona Press) pp~327--366

\item[] Menou K 2000 {\it Science} {\bf 288} 2022--2024

\item[] Mercer-Smith J A, Cameron A G W and Epstein R I 1984 {\it
  Astrophys.\ J.} {\bf 279} 363--366

\item[] Mestel L 1965a {it Quart.\ J. R. Astron.\ Soc.} {\bf 6} 161--198

\item[] Mestel L 1965b {it Quart.\ J. R. Astron.\ Soc.} {\bf 6} 265--298

\item[] Mestel L and Spitzer L 1956 {\it Mon.\ Not.\ R. Astron.\ Soc.}
  {\bf 116} 503--514

\item[] Meyer M R, Adams F C, Hillenbrand L A, Carpenter J M and Larson
  R B 2000 {\it Protostars and Planets IV} ed V Mannings \etal (Tucson:
  Univ.\ of Arizona Press) pp~121--149

\item[] Miller G E and Scalo J M 1979 {\it Astrophys.\ J. Suppl.}
  {\bf 41} 513--547

\item[] Miyama S M, Hayashi C and Narita S 1984 {\it Astrophys.\ J.}
  {\bf 279} 621--632

\item[] Miyama S M, Narita S and Hayashi C 1987 {\it Prog.\ Theor.\
  Phys.} {\bf 78} 1273--1287

\item[] Molinari S, Testi L, Rodriguez L F and Zhang Q 2002 {\it
  Astrophys.\ J.} {\bf 570} 758--778

\item[] Monaghan J~J and Lattanzio J C 1991 {\it Astrophys.\ J.}
  {\bf 375} 177--189

\item[] Motte F and Andr\'e P 2001a {\it From Darkness to Light: Origin
  and Evolution of Young Stellar Clusters (ASP Conf.\ Ser.\ 243)}
  ed T Montmerle and P Andr\'e (San Francisco: Astron.\ Soc.\ Pacific)
  pp~301--312

\item[] Motte F and Andr\'e P 2001b {\it Astron.\ Astrophys.} {\bf 365}
  440--464

\item[] Motte F, Andr\'e P and Neri R 1998 {\it Astron.\ Astrophys.}
  {\bf 336} 150--172

\item[] Mouschovias T Ch 1977 {\it Astrophys.\ J.} {\bf 211} 147--151

\item[] Mouschovias T Ch 1990 {\it Physical Processes in Fragmentation
  and Star Formation} ed R Capuzzo-Dolcetta \etal (Dordrecht:\ Kluwer)
  pp~117--148

\item[] Mouschovias T Ch 1991 {\it The Physics of Star Formation and
  Early Stellar Evolution} ed C~J Lada and N~D Kylafis (Dordrecht:
  Kluwer) pp~61--122

\item[] Mouschovias T Ch and Ciolek G E 1999 {\it The Origin of Stars
  and Planetary Systems} ed C~J Lada and N~D Kylafis (Dordrecht:
  Kluwer) pp~305--339

\item[] Mundy L G, Looney L W and Welch W J 2000 {\it Protostars and
  Planets IV} ed V Mannings \etal (Tucson:\ Univ.\ of Arizona Press)
  pp~355--376

\item[] Myers P C 1983 {\it Astrophys.\ J.} {\bf 270} 105--118

\item[] Myers P C 1985 {\it Protostars and Planets II} ed D~C Black
  and M~S Matthews (Tucson:\ Univ.\ of Arizona Press) pp~81--103

\item[] Myers P C 1999 {\it The Origin of Stars and Planetary Systems}
  ed C~J Lada and N~D Kylafis (Dordrecht:\ Kluwer) pp~67--95

\item[] Myers P C and Goodman A~A 1988 {\it Astrophys.\ J.} {\bf 329}
  392--405

\item[] Myers P C and Lazarian A 1998 {\it Astrophys.\ J.} {\bf 507}
  L157--160

\item[] Myers P C, Evans N J and Ohashi N 2000 {\it Protostars and
  Planets IV} ed V Mannings \etal (Tucson:\ Univ.\ of Arizona Press)
  pp~217--245

\item[] Nakamura T 1984 {\it Prog.\ Theor.\ Phys.} {\bf 71} 212--214

\item[] Nakamura F 2000 {\it Astrophys.\ J.} {\bf 543} 291--298

\item[] Nakamura F. Hanawa T and Nakano T 1993 {\it Publ.\ Astron.\
  Soc.\ Japan} {\bf 45} 551--566

\item[] Nakamura F, Hanawa T and Nakano T 1995 {\it Astrophys.\ J.}
  {\bf 444} 770--786

\item[] Nakamura F, Matsumoto T, Hanawa T and Tomisaka K 1999 {\it
  Astrophys.\ J.} {\bf 510} 274--290

\item[] Nakano T 1984 {\it Fundam.\ Cosmic Phys.} {\bf 9} 139--232

\item[] Nakano T 1989 {\it Astrophys.\ J.} {\bf 345} 464--471

\item[] Nakano T 1998 {\it Astrophys.\ J.} {\bf 494} 587--604

\item[] Nakano T and Nakamura T 1978 {\it Publ.\ Astron.\ Soc.\ Japan}
  {\bf 30} 671--680

\item[] Nakano T, Hasegawa T and Norman C 1995 {\it Astrophys.\ J.}
  {\bf 450} 183--195

\item[] Narita S, Hayashi C and Miyama S M 1984 {\it Prog.\ Theor.\
  Phys.} {\bf 72} 1118--1136

\item[] Nelson, A F 2000 {\it Birth and Evolution of Binary Stars
  (Poster Proceedings of IAU Symp.\ 200)} ed B Reipurth and H Zinnecker
  (Potsdam: Astrophys.\ Inst.\ Potsdam) pp~205--207

\item[] Nomura H and Mineshige S 2000 {\it Astrophys.\ J.} {\bf 536}
  429--437

\item[] Nordlund A and Padoan P 2003 {\it Turbulence and Magnetic Fields
  Astrophysics (Lecture Notes in Physics 614)} ed E Falgarone and
  T Passot (Berlin: Springer) pp~271--298

\item[] Norman M L and Wilson J R 1978 {\it Astrophys.\ J.} {\bf 224}
  497--511

\item[] Norman M L, Wilson J R and Barton R T 1980 {\it Astrophys.\ J.}
  {\bf 239} 968--981

\item[] Ogino S, Tomisaka K and Nakamura F 1999 {\it Publ.\ Astron.\
  Soc.\ Japan} {\bf 51} 637--651

\item[] Ohashi N 1999 {\it Star Formation 1999} ed T Nakamoto
  (Nobeyama: Nobeyama Radio Observatory) pp~129--135

\item[] Ohashi N, Hayashi M, Ho P T P, Momose M, Tamura M, Hirano N
  and Sargent A I 1997 {\it Astrophys.\ J.} {\bf 488} 317--329

\item[] Ostriker E C, Gammie C F and Stone J M 1999 {\it Astrophys.\
  J.} {\bf 513} 259--274

\item[] Ostriker E C, Stone J M and Gammie C F 2001 {\it Astrophys.\
  J.} {\bf 546} 980--1005

\item[] Ouyed R and Pudritz R E 1999 {\it Mon.\ Not.\ R. Astron.\ Soc.}
  {\bf 309} 233--244

\item[] Pacholczyk A G 1963 {\it Acta Astronomica} {\bf 13} 1--29

\item[] Padoan P and Nordlund A 1999 {\it Astrophys.\ J.} {\bf 526}
  279--294

\item[] Padoan P, Juvela M, Goodman A~A and Nordlund A 2001 {\it
  Astrophys.\ J.} {\bf 553} 227--234

\item[] Palla F 1999 {\it The Origin of Stars and Planetary Systems}
  ed C~J Lada and N~D Kylafis (Dordrecht:\ Kluwer) pp~375--408

\item[] Palla F 2001 {\it From Darkness to Light: Origin and Evolution
  of Young Stellar Clusters (ASP Conf.\ Ser.\ 243)} ed T Montmerle
  and P Andr\'e (San Francisco: Astron.\ Soc.\ Pacific) pp~525--536

\item[] Palla F 2002 {\it Physics of Star Formation in Galaxies} ed
  A Maeder and G Meynet (Berlin: Springer) pp~9--133

\item[] Palla F and Stahler S W 2002 {\it Astrophys.\ J.} {\bf 581}
  1194--1203

\item[] Papaloizou J C B and Lin D N C 1995 {\it Ann.\ Rev.\ Astron.\
  Astrophys.} {\bf 33} 505--540

\item[] Penston M V 1966 {\it Royal Observatory Bulletins, Series E,
  No.} {\bf 117} 299--312

\item[] Penston M V 1969 {\it Mon.\ Not.\ R. Astron.\ Soc.} {\bf 144}
  425--448

\item[] Portegies Zwart S F and McMillan S L W 2002 {\it Astrophys.\ J.}
  {\bf 576} 899--907

\item[] Prasad S~S, Tarafdar S P, Villere K R and Huntress W T 1987
  {\it Interstellar Processes} ed D~J Hollenbach and H~A Thronson
  (Dordrecht:\ Reidel) pp~631--666

\item[] Preibisch T, Weigelt G and Zinnecker H 2001 {\it The Formation
  of Binary Stars (IAU Symp.\ 200)} ed H Zinnecker and R~D Mathieu (San
  Francisco: Astron.\ Soc.\ Pacific) pp~69--78

\item[] Quataert E and Chiang E I 2000 {\it Astrophys.\ J.} {\bf 543}
  432--437

\item[] Reipurth B 1989 {\it Nature} {\bf 340} 42--45

\item[] Reipurth B 1991 {\it The Physics of Star Formation and Early
  Evolution} ed C~J Lada and N~D Kylafis (Dordrecht:\ Kluwer)
  pp~497--538

\item[] Reipurth B 2000 {\it Astron.\ J.} {\bf 120} 3177--3191

\item[] Reipurth B 2001 {\it The Formation of Binary Stars (IAU Symp.\
  200)} ed H Zinnecker and R~D Mathieu (San Francisco: Astron.\ Soc.\
  Pacific) pp~249--260

\item[] Rice W K M, Armitage P J, Bate M R and Bonnell I A 2003
  {\it Mon.\ Not.\ R. Astron.\ Soc.} {\bf 339} 1025--1030

\item[] R\'o\.zyczka M and Spruit H C 1993 {\it Astrophys.\ J.}
  {\bf 417} 677--86

\item[] Saigo K and Hanawa T 1998 {\it Astrophys.\ J.} {\bf 493}
  342--350

\item[] Saigo K, Matsumoto T and Hanawa T 2000 {\it Astrophys.\ J.}
  {\bf 531} 971-987

\item[] Salpeter E~E 1955 {\it Astrophys.\ J.} {\bf 121} 161--167

\item[] Savonije G J, Papaloizou J C B and Lin D N C 1994 {\it Mon.\
  Not.\ R. Astron.\ Soc.} {\bf 268} 13--28

\item[] Sawada K, Matsuda T, Inoue M and Hachisu I 1987 {\it Mon.\
  Not.\ R. Astron.\ Soc.} {\bf 224} 307--322

\item[] Scalo J M 1985 {\it Protostars and Planets II} ed D~C Black
  and M~S Matthews (Tucson:\ Univ.\ of Arizona Press) pp~201--296

\item[] Scalo J M 1986 {\it Fundam.\ Cosmic Phys.} {\bf 11} 1--278

\item[] Scalo J M 1987 {\it Interstellar Processes} ed D~J Hollenbach
  and H~A Thronson (Dordrecht:\ Reidel) pp~349--392

\item[] Scalo J M 1990 {\it Physical Processes in Fragmentation and
  Star Formation} ed R Capuzzo-Dolcetta \etal (Dordrecht:\ Kluwer)
  p~151--177

\item[] Scalo J M 1998 {\it The Stellar Initial Mass Function (ASP
  Conf.\ Ser.\ 142)} ed G Gilmore and D Howell (San Francisco:
  Astron.\ Soc.\ Pacific) pp~201--241

\item[] Schneider S and Elmegreen B G 1979 {\it Astrophys.\ J. Suppl.}
  {\bf 41} 87--95

\item[] Shakura N I and Sunyaev R A 1973 {\it Astron.\ Astrophys.}
  {\bf 24} 337--355

\item[] Shu F H 1976 {\it Structure and Evolution of Close Binary
  Systems (IAU Symp.\ 73)} ed P Eggleton, S Mitton and J Whelan
  (Dordrecht: Reidel) pp~ 253--264

\item[] Shu F H 1977 {\it Astrophys.\ J.} {\bf 214} 488--497

\item[] Shu F H, Adams F C and Lizano S 1987 {\it Ann.\ Rev.\ Astron.\
  Astrophys.} {\bf 25} 23--81

\item[] Shu F, Najita J, Galli D, Ostriker E and Lizano S 1993 {\it
  Protostars and Planets III} ed E~H Levy and J~I Lunine (Tucson:\
  Univ.\ of Arizona Press) pp~3--45

\item[] Shu F H, Allen A, Shang H, Ostriker E C and Li Z-Y 1999 {\it
  The Origin of Stars and Planetary Systems} ed C~J Lada and N~D
  Kylafis (Dordrecht:\ Kluwer) pp~193--226

\item[] Shu F H, Najita J R, Shang H and Li Z-Y 2000 {\it Protostars
  and Planets IV} ed V Mannings \etal (Tucson:\ Univ.\ of Arizona
  Press) pp~789--813

\item[] Sigalotti L Di G and Klapp J 2001 {\it Intl.\ J. Mod.\ Phys.\ D}
  {\bf 10} 115--211

\item[] Simon R 1965 {\it Annales d'Astrophysique} {\bf 28} 40--45

\item[] Simon M 1997 {\it Astrophys.\ J.} {\bf 482} L81--84

\item[] Simon M, Ghez A M, Leinert Ch, Cassar L, Chen W P, Howell R~R,
  Jameson R F, Matthews K, Neugebauer G, and Richichi A 1995 {\it
  Astrophys.\ J.} {\bf 443} 625--637

\item[] Solomon P M and Sanders D B 1985 {\it Protostars and Planets II}
  ed D~C Black and M~S Matthews (Tucson:\ Univ.\ of Arizona Press)
  pp~59--80

\item[] Spitzer L 1942 {\it Astrophys.\ J.} {\bf 95} 329--344

\item[] Spitzer L 1968 {\it Nebulae and Interstellar Matter} ed B~M
  Middlehurst and L~H Aller (Chicago: Univ.\ of Chicago Press) pp~1--63

\item[] Spitzer L 1978 {\it Physical Processes in the Interstellar
  Medium} (New York: Wiley-Interscience)

\item[] Spruit H C 1987 {\it Astron.\ Astrophys.} {\bf 184} 173--184

\item[] Spruit H C 1991 {\it Rev.\ Mod.\ Astron.} {\bf 4} 197--207

\item[] Spruit H C, Matsuda T, Inoue M and Sawada K 1987 {\it Mon.\
  Not.\ R. Astron.\ Soc.} {\bf 229} 517--527 

\item[] Stahler S W 1983 {\it Astrophys.\ J.} {\bf 274} 822--829

\item[] Stahler S W 1984 {\it Astrophys.\ J.} {\bf 281} 209--218

\item[] Stahler S W 1988 {\it Astrophys.\ J.} {\bf 332} 804--825

\item[] Stahler S W 1994 {\it Publ.\ Astron.\ Soc. Pacific} {\bf 106}
  337--343

\item[] Stahler S W 2000 {\it Star Formation from the Small to the
  Large Scale} ed F Favata \etal (Noordwijk: ESA Publications,
  ESA SP-445) pp~133--140

\item[] Stahler S W and Walter F M 1993 {\it Protostars and Planets
  III} ed E~H Levy and J~I Lunine (Tucson:\ Univ.\ of Arizona Press)
  pp~405--428

\item[] Stahler S W, Shu F H and Taam R E 1980a {\it Astrophys.\ J.}
  {\bf 241} 637-6-54

\item[] Stahler S W, Shu F H and Taam R E 1980b {\it Astrophys.\ J.}
  {\bf 242} 226--241

\item[] Stahler S W, Shu F H and Taam R E 1981 {\it Astrophys.\ J.}
  {\bf 248} 727--737

\item[] Stahler S W, Korycanski D G, Brothers M J and Touma J 1994
  {\it Astrophys.\ J.} {\bf 431} 341--358

\item[] Stahler S W, Palla F and Ho P T P 2000 {\it Protostars and
  Planets IV} ed V Mannings \etal (Tucson:\ Univ.\ of Arizona Press)
  pp~327--351

\item[] Sterzik M F and Durisen R H 1998 {\it Astron.\ Astrophys.}
  {\bf 339} 95--112

\item[] Sterzik M F and Durisen R H 1999 {\it Star Formation 1999}
  ed T Nakamoto (Nobeyama: Nobeyama Radio Observatory) pp~387--388

\item[] Stodolkiewicz J S 1963 {\it Acta Astronomica} {\bf 13} 30--54

\item[] Stone J M, Ostriker E C and Gammie C F 1998 {\it Astrophys.\ J.}
  {\bf 508} L99--102

\item[] Stone J M, Gammie C F, Balbus S A and Hawley J F 2000 {\it
  Protostars and Planets IV} ed V Mannings \etal (Tucson:\ Univ.\ of
  Arizona Press) pp~589--611

\item[] Stutzki J, Bensch F, Heithausen A, Ossenkopf V and Zielinsky M
  1998 {\it Astron.\ Astrophys.} {\bf 336} 697--720

\item[] Susa H and Nakamoto T 2002 {\it Astrophys.\ J.} {\bf 564}
  L57--L60

\item[] Tenorio-Tagle G 1979  {\it Astron.\ Astrophys.} {\bf 71} 59--65

\item[] Tenorio-Tagle G and Bodenheimer P 1988 {\it Ann.\ Rev.\ Astron.\
  Astrophys.} {\bf 26} 145--197

\item[] Tenorio-Tagle G, Prieto M and S\'anchez F 1992 (eds) {\it Star
  Formation in Stellar Systems} (Cambridge: Cambridge Univ.\ Press)

\item[] Terebey S, Shu F H and Cassen P 1984 {\it Astrophys.\ J.}
  {\bf 286} 529--551

\item[] Testi L and Sargent A I 1998 {\it Astrophys.\ J.} {\bf 508}
  L91--L94

\item[] Testi L, Palla F and Natta A 1999 {\it Astron.\ Astrophys.}
  {\bf 342} 515--523

\item[] Testi L, Sargent A I, Olmi L and Onello J S 2000 {\it Astrophys.\
  J.} {\bf 540} L53--56

\item[] Testi L, Palla F and Natta A 2001 {\it From Darkness to Light:
  Origin and Evolution of Young Stellar Clusters (ASP Conf.\ Ser.\
  243)} ed T Montmerle and P Andr\'e (San Francisco: Astron.\ Soc.\
  Pacific) pp~377--386

\item[] Tohline J E 1980 {\it Astrophys.\ J.} {\bf 236} 160--171

\item[] Tohline J E 1982 {\it Fundam.\ Cosmic Phys.} {\bf 8} 1--81

\item[] Tomisaka K 1996a {\it Publ.\ Astron.\ Soc.\ Japan} {\bf 48}
  701--717

\item[] Tomisaka K 1996b {\it Publ.\ Astron.\ Soc.\ Japan} {\bf 48}
  L97--L101

\item[] Tomisaka K 2002 {\it Astrophys.\ J.} {\bf 575} 306--326

\item[] Toomre A 1981 {\it The Structure and Evolution of Normal
  Galaxies} ed S~M Fall and D Lynden-Bell (Cambridge: Cambridge Univ.\
  Press) pp~111--136

\item[] Toomre A 1982 {\it Astrophys.\ J.} {\it bf 259} 535--543

\item[] Tsuribe T and Inutsuka S 1999a {\it Astrophys.\ J.} {\bf 523}
  L155--L158

\item[] Tsuribe T and Inutsuka S 1999b {\it Astrophys.\ J.} {\bf 526}
  307--313

\item[] van der Marel R P 2003 {\it Coevolution of Black Holes and
  Galaxies (Carnegie Observatories Astrophysics Series, Vol.\ 1)} ed
  L~C Ho (Cambridge: Cambridge Univ.\ Press) in press (astro-ph/0302101)

\item[] van Dishoeck E F and Blake G A 1998 {\it Ann.\ Rev.\ Astron.\
  Astrophys.} {\bf 36} 317--368

\item[] V\'azquez-Semadeni E, Ostriker E C, Passot T, Gammie C F and
  Stone J M 2000 {\it Protostars and Planets IV} ed V Mannings \etal
  (Tucson:\ Univ.\ of Arizona Press) pp~3--28

\item[] Visser A E, Richer J S and Chandler C J 2002 {\it Astron.\ J.}
  {\bf 124} 2756--2789

\item[] Ward-Thompson D 2002 {\it Science} {\bf 295} 76--81

\item[] Whitworth A 2001 {\it The Formation of Binary Stars (IAU Symp.\
  200)} ed H Zinnecker and R~D Mathieu (San Francisco: Astron.\ Soc.\
  Pacific) pp~33--39

\item[] Whitworth A and Summers D 1985 {\it Mon.\ Not.\ R. Astron.\ 
  Soc.} {\bf 214} 1--25

\item[] Whitworth A P, Bhattal A S, Francis N and Watkins S J 1996
  {\it Mon.\ Not.\ R. Astron.\ Soc.} {\bf 283} 1061--1070

\item[] Whitworth A P, Boffin H M J and Francis N 1998 {\it Mon.\ Not.\
  R. Astron.\ Soc.} {\bf 299} 554--561

\item[] Williams J P and Myers P C 2000 {\it Astrophys.\ J.} {\bf 537}
  891--903

\item[] Williams J P, Blitz L and McKee C F 2000 {\it Protostars and
  Planets IV} ed V Mannings \etal (Tucson:\ Univ.\ of Arizona Press)
  pp~97--120

\item[] Winkler K-H A and Newman M J 1980a {\it Astrophys.\ J.}
  {\bf 236} 201--211

\item[] Winkler K-H A and Newman M J 1980b {\it Astrophys.\ J.}
  {\bf 238} 311--325

\item[] Wolfire M G and Cassinelli J P 1987 {\it Astrophys.\ J.}
  {\bf 319} 850--867

\item[] Wuchterl G and Tscharnuter W M 2003 {\it Astron.\ Astrophys.}
  {\bf 398} 1081--1090

\item[] Yorke H W and Bodenheimer P 1999 {\it Astrophys.\ J.} {\bf 525}
  330--342

\item[] Yorke H W and Kr\"ugel E 1977 {\it Astron.\ Astrophys.} {\bf 54} 
  183--194

\item[] Yorke H W and Sonnhalter C 2002 {\it Astrophys.\ J.} {\bf 569}
  846--862

\item[] Zinnecker H 1982 {\it Symposium on the Orion Nebula to Honor
  Henry Draper} ed A~E Glassgold \etal (New York: Ann.\ New York
  Acad.\ Sci.\ Vol.\ 395) pp~226--235

\item[] Zinnecker H 1984 {\it Mon.\ Not.\ R. Astron.\ Soc.}
  {\bf 210} 43--56

\item[] Zinnecker H and Bate M R 2002 {\it Hot Star Workshop III: The
  Earliest Stages of Massive Star Birth (ASP Conf.\ Ser.\ 267)}
  ed P A Crowther (San Francisco: Astron.\ Soc.\ Pacific) pp~209--218

\item[] Zinnecker H and Mathieu R D 2001 (eds) {\it The Formation of
  Binary Stars (IAU Symp.\ 200)} (San Francisco: Astron.\ Soc.\ Pacific)

\item[] Zinnecker H, McCaughrean M J and Wilking B A 1993 {\it
  Protostars and Planets III} ed E~H Levy and J~I Lunine (Tucson:
  Univ.\ of Arizona Press) pp~429--495

\endrefs

\end{document}